\documentclass[a4paper,fleqn,usenatbib]{mnras}

\usepackage{newtxtext,newtxmath}
\usepackage[T1]{fontenc}
\usepackage{ae,aecompl}

\usepackage{graphicx}	% Including figure files
\usepackage{amsmath}	% Advanced maths commands
\usepackage{amssymb}	% Extra maths symbols

\title[Grain size distribution in galaxies]{Remodeling the evolution of grain size distribution in galaxies}

\author[H. Hirashita and S. Aoyama]{
Hiroyuki Hirashita\thanks{E-mail: hirashita@asiaa.sinica.edu.tw}
and Shohei Aoyama %%$^1$
\\
Institute of Astronomy and Astrophysics, Academia Sinica,
Astronomy-Mathematics Building, AS/NTU
No.\ 1, Sec.\ 4, Roosevelt Road,\\ Taipei 10617, Taiwan %%\\
}

\date{Accepted XXX. Received YYY; in original form ZZZ}

\pubyear{2018}

\begin{document}
\label{firstpage}
\pagerange{\pageref{firstpage}--\pageref{lastpage}}
\maketitle

\begin{abstract}
We revisit the evolution model of grain size distribution
in a galaxy for the ultimate purpose of implementing it
in hydrodynamical simulations.
We simplify the previous model in
such a way that some model-dependent assumptions are
replaced with simpler functional forms. For the first test of
the developed framework, we apply it to
a one-zone chemical evolution model of a galaxy, confirming
that our new model satisfactorily reproduces the
previous results and that efficient coagulation
of small grains produced by shattering and accretion
is essential in reproducing the so-called MRN grain size distribution.
For the next step, in order to test if our model can be treated together with the
hydrodynamical evolution of the interstellar medium (ISM), we
post-process a hydrodynamical simulation of an isolated disc
galaxy using the new grain evolution model.  We sample
hydrodynamical particles representing each of the dense and diffuse ISM phases.
By this post-processing, we find that the processes occurring in the dense
gas (grain growth by accretion and coagulation) are important in reproducing the
grain size distribution consistent with the Milky Way extinction curve.
In our model, the grain size distributions are similar between the dense and diffuse
ISM, although we observe a larger dispersion in the dense ISM.
Moreover, we also show that even if we degrade the grain radius resolution
(with 16 grid points), the overall shape of grain size distribution
(and of resulting extinction curve) can be captured.
\end{abstract}

\begin{keywords}
dust, extinction --- galaxies: evolution
--- galaxies: ISM --- galaxies: spiral --- methods: numerical
\end{keywords}

%%%%%%%%%%%%%%%%%%%%%%%%%%%%%%%%%%%%%%%%%%%%%%%%%%

%%%%%%%%%%%%%%%%% BODY OF PAPER %%%%%%%%%%%%%%%%%%

\section{Introduction}

The evolution of dust in the interstellar medium (ISM) plays a fundamental role in
characterizing the galaxy evolution.
Dust modifies the appearance of galaxies by absorbing and
scattering the stellar light
and reemitting it into the far-infrared (FIR).\footnote{In this paper, we use
the term FIR for the wavelength range where the emission is dominated by
dust.} Thus, dust shapes the
observed spectral energy distributions (SEDs) of galaxies from
ultraviolet (UV) to FIR \citep[e.g.][]{Takeuchi:2005aa}.
Dust surfaces are
the main site for the formation of some molecular species, especially
H$_2$ \citep[e.g.][]{Gould:1963aa,Cazaux:2004aa}, giving rise to molecular-rich
environments or molecular clouds, which host star formation
\citep[e.g.][]{Hirashita:2002aa,Chen:2018aa}.
Dust also affects the thermal evolution of star-forming clouds by radiating
away the thermal energy (i.e.\ dust cooling).
In the later stage of star formation, dust cooling induces fragmentation
\citep{Omukai:2005aa} and determines the typical stellar mass
\citep{Schneider:2006aa}.

The above important processes are not only scaled with the total dust
abundance but also affected by the grain size distribution.
In particular, the grain size distribution is important in determining the
efficiencies of grain surface reaction rates \citep{Yamasawa:2011aa,Harada:2017aa}
and the wavelength dependence
of dust opacity \citep[e.g.][hereafter MRN]{Mathis:1977aa}.
Therefore, we have to understand the evolutions of both total dust
abundance and grain size distribution. The total dust abundance is
predominantly determined by stellar dust production, dust growth
(accretion of gas-phase materials),
and dust destruction [sputtering in supernova (SN) shocks]
\citep[e.g.][]{Dwek:1998aa,Zhukovska:2008aa},
while the grain size distribution is determined by
shattering (grain fragmentation) and coagulation (grain--grain sticking)
in addition to the above processes \citep[e.g.][hereafter A13]{Asano:2013aa}.
In fact, as pointed out by \citet{Kuo:2012aa}, the total dust abundance is
also affected by the grain size distribution because grain growth by accretion occurs
in a way dependent on the total grain surface area. This indicates that the dust evolution
model should include the two aspects (total dust abundance and grain
size distribution) in a consistent manner.

Recently, there have been some efforts of consistently modeling the evolution
of total dust abundance and grain size distribution. A13
constructed a full framework
for treating the evolution of grain size
distribution throughout the entire galactic evolution. After such an effort,
the evolution of grain size distribution is found to be roughly
described in the following way.
At the early metal-poor stage of galactic evolution, the dust is
predominantly supplied from SNe
\citep{Kozasa:1989aa,Todini:2001aa,Nozawa:2003aa,Bianchi:2007aa,Cherchneff:2010aa}
and asymptotic giant branch (AGB) stars
\citep[e.g.][]{Ferrarotti:2006aa,Valiante:2009aa,Ventura:2014aa}.
In this phase, shattering gradually produces small grains
\citep[e.g.][]{Hirashita:2012aa}. When the system is enriched with metals,
small grains produced by shattering start to accrete a significant
amount of gas-phase metals in the dense ISM because small grains
have large surface-to-volume ratios compared with large grains
\citep{Kuo:2012aa}. As a consequence,
the abundance of small grains drastically increases. This dust growth
mechanism (accretion) dominates the dust mass in galaxies whose
metallicity is higher than $\sim$0.1--0.3 Z$_{\sun}$
\citep[e.g.][]{Dwek:1998aa,Zhukovska:2008aa}.
Afterwards, coagulation in dense clouds converts small grains to large
grains. Therefore, the small grain abundance
is the most enhanced relative to the total dust abundance at
$\sim$0.3 Z$_{\sun}$, which means that the extinction curve is the
steepest at sub-solar metallicity \citep{Hou:2017aa}.

Dust evolution models have recently been included in hydrodynamical
simulations of galaxies for the purpose of obtaining a consistent understanding
between the dust and ISM evolution. 
Although dust could be roughly modeled by assuming a tight relation with
metallicity \citep{Yajima:2014aa},
some of the processes
driving dust evolution (accretion, coagulation, shattering, and destruction) are
affected by the density and temperature of the ISM as well as the metallicity.
Therefore, it is
desirable to model dust and the ISM consistently. Hydrodynamical
simulations provide viable
methods of computing the evolution of the ISM.
There have been some studies incorporating dust evolution in galaxy-scale
or cosmological hydrodynamical simulations.
\citet{Dayal:2010aa} computed dust production and destruction in a
cosmological simulation and predicted the FIR luminosities of high-redshift
galaxies based on their models of
Lyman $\alpha$ emitters and Lyman break galaxies.
\citet{Bekki:2015ab} treated dust as a separate component
and calculated the dust evolution under the interaction with
gas, stars and dark matter.
As an application, they also incorporated H$_2$ formation on dust surfaces
in a consistent manner with dust abundance evolution to investigate
the spatial distribution of dust and molecular gas in galaxies
\citep[see also][]{Bekki:2013aa}.
\citet{McKinnon:2016aa} solved dust evolution in cosmological zoom-in simulations.
They broadly reproduced the relation between dust-to-gas ratio and various quantities
in galaxies, but they still needed a more realistic treatment of dust destruction
and feedback by SNe. \citet{McKinnon:2017aa} extended their model to full-volume
cosmological predictions, which are useful to examine statistical properties of dust,
especially, the dust mass function and the comoving dust mass density. Their
simulation broadly reproduced the relevant observations in
the present-day Universe. \citet{Zhukovska:2016aa} analyzed the effect of dust growth
by accretion on the relation between gas density and depletion (or dust-to-metal ratio)
in an isolated
Milky Way-like galaxy by post-processing a hydrodynamical simulation. They
examined gas-temperature-dependent sticking coefficient in accretion,
in order to reproduce the relation between silicon depletion and gas density.
There are also some semi-analytic models for dust evolution in a cosmological volume,
focusing on high-redshift galaxies \citep{Valiante:2011aa,Mancini:2015aa}
or including the comparison with the local
Universe \citep{Popping:2017aa,Ginolfi:2018aa}.

None of the above simulations treated the grain size distribution.
As mentioned above, the grain size distribution
affects the dust evolution. Implementation of grain size distributions in a
galaxy-scale or cosmological hydrodynamical simulation has not been
successful (except for the recent preliminary runs in \citealt{McKinnon:2018aa}),
mainly because of the high computational cost.
For the purpose of treating the evolution of grain size distribution within the
limitation of available computational capability, \citet[][hereafter A17]{Aoyama:2017aa}
and \citet{Hou:2017aa} represented the grain sizes by
two sizes ranges divided at around $a\sim 0.03~\micron$ ($a$ is the grain radius),
according to the formulation by \citet{Hirashita:2015aa}.
As a consequence of this two-size approximation,
they succeeded in computing the spatial variations
not only in the dust abundance, but also in the grain size distribution.
In particular, they predicted
the spatial and temporal variation in
extinction curves based on the calculated grain size distributions.
The two-size approximation has been applied to a simulation of
clusters of galaxies by \citet{Gjergo:2018aa}. \citet{Aoyama:2018aa}
have recently extended the simulation to a cosmological volume,
predicting the statistical properties of dust abundance in galaxies and
in the intergalactic medium (IGM).
They confirmed the evolution of grain size distribution predicted in
A13's one-zone model, but at the same time, they showed that the
grain size distribution has a large variety among galaxies with similar
mass and metallicity (see also Hou et al., in preparation).
Such a variation cannot be naturally treated by one-zone models.
Moreover, the grain size distribution in the IGM can only be predicted by
cosmological simulations because the IGM dust is supplied from a
large number of galaxies by stellar feedback \citep[see also][]{Zu:2011aa}.
Since the observational estimate of dust mass in the IGM depends on the grain size
distribution \citep{Menard:2012aa}, it is desirable to clarify using cosmological
simulations how the grain size
distribution evolves in the IGM.

Although the above two-size approximation significantly reduces the computational cost, 
the prediction of extinction curves under such a treatment
depends on how we reconstruct the grain size distribution from
the two-size information. In other words, since
the entire grain size range is represented by two sizes, the precision of
the predicted extinction curves is limited.
{Other methods that avoid treating the full grain size distribution
such as the moment formulation by \citet{Mattsson:2016aa} also have the
same problem.} More sampling of grain sizes
increases the predictive power of extinction curves.
For more improvement of the above theoretical predictions, thus, it is
a natural step to directly treat the grain size distribution without
the two-size approximation or {the moment formulation} in galactic or cosmological simulations.
\citet{McKinnon:2018aa} have recently developed a full grain size distribution
calculation and given some test cases for isolated galaxy simulations,
although realistic predictions still need to wait for their future work.

There are some possibilities of reducing the computational cost
in calculations of grain size distribution.
In A13's model, there are some components
that depend on specific assumptions or models. For example, they adopted the results
of \citet{Nozawa:2006aa} for dust destruction in SN shocks, but this
procedure consumes computational time and memory because it
needs to read a large table for their function $\xi (a,\, a')$, which describes
the transition probability of the grain radius from $a'$ to $a$.
However, as shown by \citet{Nozawa:2006aa}, the efficiency of
SN destruction depends on the ambient gas density;
this dependence causes uncertainties in estimating the dust destruction rate,
since each SN is hardly resolved in galaxy-scale simulations.
For shattering and coagulation, A13
adopted specific calculations for grain velocities from \citet{Yan:2004aa};
however, the grain velocities depend on the ambient physical conditions
such as gas density, magnetic field strength, ionization degree, etc.
It is extremely time-consuming to consider such an environmental dependence,
and most hydrodynamical simulations on galaxy scales do not have
predictive power of all the relevant physical
quantities because of limited spatial resolutions.
In the treatment of shattered fragments, A13 is based on
\citet{Jones:1996aa}, whose formulation contains a few material parameters,
but \citet{Hirashita:2013ab} show that a simpler
treatment by \citet{Kobayashi:2010aa}, who basically describe shattering
with a single parameter (critical velocity for catastrophic disruption),
produces satisfactorily similar results. Simplifying some of the above assumptions
will save the computational cost, but still will not change the results within the modeling
uncertainties.

Given the above possible simplifications, it is worth remodeling
the evolution of grain size distribution.
This is the first purpose of this paper. The second aim is
to test if the new model gives consistent results with the previous
models such as A13's.
The third is to investigate a possibility of implementing the new formalism
in a hydrodynamic simulation. To this goal, we post-process A17's
simulation of an isolated disc galaxy to derive the evolution of grain size distribution in
some selected fluid elements of the ISM. This step gives us an idea about
not only the suitability for the implementation into hydrodynamical
codes but also the dependence of the grain size distribution
on the local physical conditions in a galaxy. We also test how the calculated
grain size distribution depends on the grain radius resolution, since minimizing
the number of grain size grids (bins) is useful to save the computational power.

This paper is organized as follows.
In Section \ref{sec:model}, we formulate the evolution of grain size distribution.
In Section \ref{sec:application}, we apply the formulation to {a} one-zone model
and compare it with some previous results (mainly with A13).
In Section \ref{sec:result}, we post-process a hydrodynamical simulation of
an isolated disc galaxy with our newly developed model.
In Section \ref{sec:discussion}, we discuss some issues and prospect
for our dust evolution model.
In Section \ref{sec:conclusion}, we give the conclusion of
this paper.
%%We use $(h,\,\Omega_\mathrm{m},\,\Omega_\Lambda )=(0.7,\, 0.3,\, 0.7)$
%%for the cosmological parameters.

\section{Basic equations}\label{sec:model}

We basically follow A13's model for the evolution of grain size distribution,
but simplify it as long as we do not lose the physical essence
of the relevant processes. We also remove some model-dependent
ingredients from A13's model, and, if necessary, replace them with
simpler formulae. This simplification could be useful
for the purpose of making the dust evolution model suitable for implementation
in hydrodynamical simulations.
For the dust evolution processes,
we consider stellar dust production, dust destruction by SN shocks in
the ISM, dust growth by accretion and coagulation in the dense ISM, and
dust disruption by shattering in the diffuse ISM.

{Since the evolution of grain size distribution is affected by
the physical condition of the ISM, we need to combine the model
(or the basic equations)
developed in this section with an appropriate evolution model of the ISM
(or a galaxy evolution model).
We adopt
two types of galaxy evolution models in Section \ref{sec:result}: one is a one-zone model and the
other is a hydrodynamical model. The one-zone model applied in
Section \ref{sec:application} has an advantage of simplicity since it neglects
the spatial structure of the ISM. Thus, this model is suitable for the first test
of the grain size evolution model newly developed in this section. The
disadvantage of the one-zone model is that we have to assume the
physical condition of the ISM. To overcome this, we use a hydrodynamical simulation
to calculate the spatially resolved structure of the ISM. We use
(or post-process)
the hydrodynamical simulation of an isolated disc galaxy to calculate the
evolution of grain size distribution. Recall that, as mentioned in the Introduction,
one of the major purposes of this paper is to develop a grain size distribution model
that can be implemented in hydrodynamical simulations.
The application in Section \ref{sec:result} will give a useful step to this goal.}

{In this section, we describe the basic equations for the evolution of
grain size distribution.}
The grain size distribution is expressed by the grain mass distribution
$\rho_\mathrm{d}(m,\, t)$, which is defined such that
$\rho_\mathrm{d}(m,\, t)\,\mathrm{d}m$
($m$ is the grain mass and $t$ is the time) is the mass density
of dust grains whose mass is between $m$ and $m+\mathrm{d}m$.
In this paper, we assume grains to be spherical and compact,
so that $m=(4\upi /3)a^3s$, where $a$ is the grain radius and $s$
is the material density of dust. If we use the grain size distribution
$n(a,\, t)$, where $n(a,\, t)\,\mathrm{d}a$ is the number density
of dust grains with radii between $a$ and $a+\mathrm{d}a$,
it is related to the above grain mass distribution as
\begin{align}
\rho_\mathrm{d}(m,\, t)\,\mathrm{d}m=\frac{4}{3}\upi a^3sn(a)\,\mathrm{d}a,
\label{eq:rho_n}
\end{align}
with $\mathrm{d}m=4\upi a^2s\,\mathrm{d}a$. The total dust mass
density $\rho_\mathrm{d,tot}(t)$ is
\begin{align}
\rho_\mathrm{d,tot}(t)=\int_0^\infty\rho_\mathrm{d}(m,\, t)\,\mathrm{d}m.
\end{align}
The gas density is given by the number density of hydrogen nuclei,
$n_\mathrm{H}$, or the gas mass density,
$\rho_\mathrm{gas}=\mu m_\mathrm{H}n_\mathrm{H}$
($\mu =1.4$ is the gas mass per hydrogen, and $m_\mathrm{H}$ is the mass
of hydrogen atom).

The time-evolution of $\rho_\mathrm{d}(m,\, t)$ is described by
\begin{align}
\frac{\upartial\rho_\mathrm{d}(m,\, t)}{\upartial t} &=
\left[\frac{\upartial\rho_\mathrm{d}(m,\, t)}{\upartial t}\right]_\mathrm{star}+
\left[\frac{\upartial\rho_\mathrm{d}(m,\, t)}{\upartial t}\right]_\mathrm{sput}
\nonumber\\
&+
\left[\frac{\upartial\rho_\mathrm{d}(m,\, t)}{\upartial t}\right]_\mathrm{acc}+
\left[\frac{\upartial\rho_\mathrm{d}(m,\, t)}{\upartial t}\right]_\mathrm{shat}
\nonumber\\
&+
\left[\frac{\upartial\rho_\mathrm{d}(m,\, t)}{\upartial t}\right]_\mathrm{coag}
+\rho_\mathrm{d}(m,\, t)\frac{\mathrm{d}\ln\rho_\mathrm{gas}}{\mathrm{d}t},
\label{eq:basic}
\end{align}
where the terms with subscripts `star', `sput', `acc', `shat' and `coag'
indicate the changing rates of grain mass distribution by stellar dust
production, sputtering, accretion, shattering, and coagulation,
respectively (those terms are evaluated below), and the last term
expresses the change of $\rho_\mathrm{d}$ caused by the change of the
background gas density (we assume that the gas and dust are dynamically
coupled) \citep{Hirashita:2015aa}. Although the equations below are
written in continuous forms, we actually solve discrete forms described in
Appendix \ref{app:discrete}. We adopt $N=128$ grid points for the
discrete grain size distribution in a grain radius range of $3\times 10^{-4}$--10 $\micron$
unless otherwise stated.
We set $\rho_\mathrm{d}(m,\, t)=0$ at the maximum and minimum grain radii
for the boundary conditions. Since the maximum radius (10 $\micron$) is large
enough, the boundary condition at the largest grain radius does not affect
the calculation. The boundary condition
at the lower boundary means that we do not regard `grains' with $a<3\times 10^{-4}~\micron$
as dust grains. The discrete time-step $\Delta t$ is chosen following
Appendix \ref{app:timestep}.
{We explain the
modeling for each process in each subsection below.}

\subsection{Stellar dust production}\label{subsec:stellar}

A certain fraction of the metals ejected from SNe and AGB stars
are condensed into dust. There are some calculations of dust
condensation in SNe \citep[e.g.][]{Kozasa:1989aa,Todini:2001aa,Nozawa:2003aa}
and AGB stars \citep[e.g.][]{Ferrarotti:2006aa,Ventura:2014aa,DellAgli:2017aa};
however, the fraction of metals eventually injected into the ISM in the form of
dust is still uncertain. For SNe, a part of condensed dust is destroyed by
the reverse shock before being injected into the ISM
\citep{Bianchi:2007aa,Nozawa:2007aa}. The destroyed fraction
depends on the ambient gas density (which is hardly resolved in
galaxy-scale simulations) and the grain size
distribution. For AGB stars, there is still a difference in the total condensed
dust mass among calculations \citep{Inoue:2011aa,Kuo:2013aa}.
Therefore (and for the sake of simplification), we choose to adopt a
constant parameter ($f_\mathrm{in}$) that describes
the condensation efficiency of metals in stellar ejecta.
For the metal enrichment, we utilize a chemical enrichment model
that outputs the metal mass injected into the ISM per unit time and
unit volume, denoted as $\dot{\rho}_{Z}$. In practice, we are usually
able to utilize a chemical enrichment model already implemented
in the hydrodynamical simulation (e.g.\ A17).
Therefore, we assume that $\dot{\rho}_Z$
as a function of time is already given.

Based on the above concept, we can write the change of the grain
size distribution by stellar dust production as
\begin{align}
\left[\frac{\upartial\rho_\mathrm{d}(m,\, t)}{\upartial t}\right]_\mathrm{star}=
f_\mathrm{in}\dot{\rho}_Z\, m\tilde{\varphi} (m),\label{eq:stellar}
\end{align}
where $m\tilde{\varphi} (m)$ is the mass distribution function of
the dust grains produced by stars, and it is normalized so that the integration
for the whole grain mass range is unity. It is often convenient to consider
the grain size distribution, so that we define
$\varphi (a)\,\mathrm{d}a\equiv\tilde{\varphi}(m)\,\mathrm{d}m$.
The typical size of stellar dust is
on the order of $\sim 0.1~\micron$; thus, we adopt the following lognormal
form for ${\varphi}(a)$:
\begin{align}
{\varphi}(a)=\frac{C_\varphi}{a}\exp\left\{
-\frac{[\ln (a/a_0)]^2}{2\sigma^2}\right\} ,
\end{align}
where $C_\varphi$ is the normalization factor, $\sigma$ is the standard deviation,
and $a_0$ is the central
grain radius. We adopt $\sigma =0.47$ and $a_0=0.1~\micron$ following A13,
who assigned these values based on the
theoretical dust condensation calculation for AGB stars by \citet{Yasuda:2012aa}.
Note that A13 separately treated SNe and AGB stars and adopted the grain
size distribution calculated by \citet{Nozawa:2007aa} for SN dust.
The normalization is determined by
\begin{align}
\int_0^\infty \frac{4}{3}\upi a^3\varphi (a)\,\mathrm{d}a =1,
\end{align}
which is equivalent to $\int m\tilde{\varphi}(m)\,\mathrm{d}m=1$.
We adopt $f_\mathrm{in}=0.1$ (A17).

\subsection{Dust destruction by SN shocks}

Since dust destruction by sputtering preserves the number of
dust grains (as long as dust is not completely destroyed),
we can apply the continuity equation in Appendix \ref{app:derivation},
obtaining \citep[see also][]{Hirashita:2015aa}
\begin{align}
\left[\frac{\upartial\rho_\mathrm{d}(m,\, t)}{\upartial t}\right]_\mathrm{sput}
=-\frac{\upartial}{\upartial t}[\dot{m}\rho_\mathrm{d}(m,\, t)]+
\frac{\dot{m}}{m}\rho_\mathrm{d}(m,\, t),\label{eq:sputtering}
\end{align}
where $\dot{m}=4\pi a^2\dot{a}$.
%%In thermal sputtering, $\dot{a}$ is independent of $a$,
%%which means that smaller grains have shorter
%%destruction time-scales \citep{Tsai:1995aa,Nozawa:2006aa}.
We estimate that
\begin{align}
%%\dot{a}=-a/\tau_\mathrm{dest}(a),
\dot{m}=-m/\tau_\mathrm{dest}(m),
\end{align}
where $\tau_\mathrm{dest}(m)$ is the destruction time-scale as
a function of grain mass.
Note that some authors adopt the decreasing time-scale of grain radius,
which is related to $\tau_\mathrm{dest}(m)$ as
$|a/\dot{a}|=3\tau_\mathrm{dest}(m)$.
We estimate $\tau_\mathrm{dest}(m)$ in what follows.

For the first step to obtain $\tau_\mathrm{dest}(m)$, we estimate the
time-scale on which the ISM is once swept by SN shocks. This is
referred to as the sweeping time-scale. The sweeping time-scale
is estimated by $M_\mathrm{gas}/(M_\mathrm{s}\gamma )$,
where $M_\mathrm{gas}$ is the gas mass of interest,
$M_\mathrm{s}$ is the gas mass swept by a single SN blast,
and $\gamma$ is the rate of SNe sweeping the gas.
Since a passage of SN shock does not destroy all the swept dust,
we introduce the destruction efficiency of dust grains,
$\epsilon_\mathrm{dest}(m)$, which is generally a function of $m$
[or $\epsilon_\mathrm{dest}(a)$, which is a function of $a$].
Using this efficiency, we estimate the destruction time-scale as a function
of grain mass as
\begin{align}
\tau_\mathrm{dest}(m)=
\frac{M_\mathrm{gas}}{\epsilon_\mathrm{dest}(m)M_\mathrm{s}\gamma}.
\label{eq:tau_dest}
\end{align}
%%where the factor 3 comes from the relation $a/\dot{a}=3m/\dot{m}$.
Considering the uncertainty in $\epsilon_\mathrm{dest}$, we
adopt an empirical value estimated by \citet{McKee:1989aa}
as $\sim 0.1$. Since the typical grain radius in the Galactic ISM is
$\sim 0.1~\micron$, we adopt $\epsilon_\mathrm{dest}=0.1$ at
$a=0.1~\micron$ and determine its dependence on $a$ below.

If we consider thermal sputtering, the destruction rate $\dot{a}$ is
independent of $a$
\citep[e.g.][]{Draine:1979aa,Tielens:1994aa,Nozawa:2006aa}.
Thus, the time-scale of destruction is proportional to $a$.
If we take relative motion between dust and gas into account,
nonthermal sputtering, which weakens the dependence of
the destruction time-scale on grain radius, occurs
\citep[e.g.][]{McKee:1987aa}.
However, small grains tend to be trapped in the shocked region,
which enhances the $a$ dependence of the destruction time-scale
\citep{Nozawa:2006aa}. These processes usually occur on unresolvable
spatial scales in galaxy-scale hydrodynamical simulations.
Therefore, for the sake of simplicity, we assume $\dot{a}$ to be
constant, or $\tau_\mathrm{dest}(m)=|a/\dot{a}|/3\propto a$.
%%Considering that the destruction efficiency is inversely
%%proportional to the destruction time-scale, we assume that
This means that
$\epsilon_\mathrm{dest}(m)=\epsilon_\mathrm{dest}(a)\propto a^{-1}$.
Recalling that
$\epsilon_\mathrm{dest}(0.1~\micron )=0.1$ as assumed above,
and that $\epsilon_\mathrm{dest}$ cannot exceed 1, we adopt
the following formula for the
destruction efficiency:\footnote{Alternatively, we can
adopt $\epsilon_\mathrm{dest}(a)=\mathrm{min}[
0.1({0.1~\micron}/{a}) ,\, 1]$, but the smooth functional form in
equation (\ref{eq:eps}) is convenient to avoid discontinuous behaviour
in numerical calculations.}
\begin{align}
\epsilon_\mathrm{dest}(a)=1-\exp\left[-0.1\left(\frac{a}{0.1~\micron}\right)\right] .
\label{eq:eps}
\end{align}
This functional form approximately realizes two asymptotic behaviours:
$\epsilon_\mathrm{dest}\sim 0.1(0.1~\micron /a)\propto a^{-1}$ at
$a\gtrsim 0.1~\micron$ and
$\epsilon_\mathrm{dest}\sim 1$ at $a\ll 0.1~\micron$.
We adopt $M_\mathrm{s}=6800$ M$_\odot$ \citep{McKee:1989aa,Nozawa:2006aa}.
The SN rate $\gamma$ is tightly coupled with the chemical enrichment;
thus, we presume a situation where $\gamma$ as a function of time is
already given (like $\dot{\rho}_Z$ above).

%%If we set an appropriate value of $\dot{a}$, we are able to
%%solve equation (\ref{eq:sputtering}). The sputtering rate in the hot
%%gas is given, e.g.\ in \citep{Tsai:1995aa} as a function of gas
%%density and temperature. In practice, the destruction only occurs
%%in the passage of SN shocks in the ISM.

%%A17 estimated the number of SN
%%shocks passing in an SPH gas particle over a small time-step $\Delta t$
%%and obtained the fraction
%%of dust destroyed in that time-step, $\eta$. By definition,
%%the destroyed dust mass density, $\Delta\rho_\mathrm{d}(a,\, t)$,
%%is expressed as
%%$\eta =\Delta\rho_\mathrm{d}(a,\, t)/\rho_\mathrm{d}(a,\, t)=\Delta a/a$.
%%Thus, $\Delta a=-\eta (a)a$. By the above assumption, $\Delta_a\equiv\eta (a)a$
%%is constant. We solve $\dot{a}=\Delta a/\Delta t=-\Delta_a/\Delta t$ over
%%$\Delta t$ (or $\dot{m}=-4\upi a^2s\Delta_a/\Delta t$ in equation \ref{eq:sputtering}).
%%Therefore, once we give $\Delta_a$ in each time-step, we obtain the
%%time evolution of grain size distribution by SN destruction.

\subsection{Dust growth by accretion}

%%The grain growth rate by accretion is independent of the grain
%%radius, i.e.\ $\dot{a}=\mbox{constant}$
%\citep{Hirashita:2011aa}. Therefore,
Since dust growth by accretion can be regarded as negative destruction, it
can be treated in a similar way to sputtering (Appendix~\ref{app:derivation}):
\begin{align}
\left[\frac{\upartial\rho_\mathrm{d}(m,\, t)}{\upartial t}\right]_\mathrm{sput}
=-\frac{\upartial}{\upartial t}[\dot{m}\rho_\mathrm{d}(m,\, t)]+
\frac{\dot{m}}{m}\rho_\mathrm{d}(m,\, t).\label{eq:accretion}
\end{align}
For accretion, the growth rate $\dot{a}$ ($\dot{m}=4\pi a^2\dot{a}$) is
estimated as
\begin{align}
\dot{m}=\xi (t)m/\tau_\mathrm{acc} (m),
\end{align}
where $\xi (t)\equiv 1-\rho_\mathrm{d,tot}(t)/\rho_Z (t)$
[$\rho_Z(t)$ is the mass density of metals in both gas and dust phases]
is the fraction of metals in the gas phase
and $\tau_\mathrm{acc} (m)$ is the grain growth time given by
\citep{Hirashita:2012aa}
\begin{align}
\tau_\mathrm{acc}(m) &= \frac{1}{3}\tau^\prime_\mathrm{acc}(a)\nonumber\\
\tau_\mathrm{acc}^\prime (a) &=  \tau^\prime_\mathrm{0,acc}\left(\frac{a}{0.1~\micron}
\right)\left(\frac{Z}{\mathrm{Z}_{\sun}}\right)^{-1}
\left(\frac{n_\mathrm{H}}{10^3~\mathrm{cm}^{-3}}
\right)^{-1}\left(\frac{T_\mathrm{gas}}{10~\mathrm{K}}
\right)^{-1/2}\nonumber\\
&\times \left(\frac{S}{0.3}\right)^{-1},
\label{eq:tau_acc}
\end{align}
where $\tau^\prime_\mathrm{0,acc}$ is a constant given below,
$Z$ is the metallicity (we adopt solar metallicity
$Z_{\sun}=0.02$ throughout this paper for the convenience
in direct comparison with previous studies), $n_\mathrm{H}$
is the hydrogen number density, $T_\mathrm{gas}$ is
the gas temperature, and $S$ is the sticking efficiency.
\citet{Hirashita:2012aa} obtain $\tau_\mathrm{0,acc}=1.61\times 10^8$
and $0.993\times 10^8$ yr for silicate and graphite, respectively.
In this paper, we apply
the silicate value unless otherwise stated.
We adopt $S=0.3$ \citep{Leitch-Devlin:1985aa,Grassi:2011aa}.
We assume a situation in which $\rho_Z$ and $n_\mathrm{H}$ are already
given; thus, the fraction of metals in the gas phase [$\xi (t)$]
and metallicity [$Z=\rho_Z/(\mu m_\mathrm{H}n_\mathrm{H})$] can be
calculated. As a
consequence, the combination of
equations (\ref{eq:accretion})--(\ref{eq:tau_acc}) is solved in
a consistent manner with the abundances of metals and dust.

\subsection{Shattering}\label{subsec:shattering}

The ISM is generally turbulent. In a turbulent medium, dust grains obtain
random velocities through the interaction with gas and magnetic
field \citep{Yan:2004aa}. In the diffuse ISM ($n_\mathrm{H}\lesssim 1$ cm$^{-3}$),
the grain velocities are so
large that shattering can occur in grain--grain collisions \citep{Hirashita:2009ab}.
We basically follow \citet{Jones:1994aa,Jones:1996aa} and \citet{Hirashita:2009ab} for the formulation
of shattering \citep[see also][]{Asano:2013aa}.

The time evolution of grain size distribution by shattering is expressed as
\begin{align}
\left[\frac{\partial\rho_\mathrm{d}(m,\, t)}{\partial t}\right]_\mathrm{shat}
= -m\rho_\mathrm{d}(m,\, t)\int_0^\infty\alpha (m_1,\, m)\rho_\mathrm{d}(m_1,\, t)
\mathrm{d}m_1\nonumber\\
+ \int_0^\infty\int_0^\infty\alpha (m_1,\, m_2)\rho_\mathrm{d}(m_1,\, t)\rho_\mathrm{d}(m_2,\, t)
\mu_\mathrm{frag}(m;\, m_1,\, m_2)\mathrm{d}m_1
\mathrm{d}m_2,\label{eq:shat}
\end{align}
where two newly introduced functions, $\alpha$ and $\mu_\mathrm{frag}$, are
explained below.
The first term on the right-hand side indicates the
loss of grains with mass $m$ by the collisions with other grains, and
$\alpha$ indicates the collision frequency normalized to the grain
masses and grain number density). If we consider collisions between
grains with masses $m_1$ and
$m_2$ (radii $a_1$ and $a_2$, respectively), $\alpha$ is expressed as
\begin{align}
\alpha (m_1,\, m_2)\equiv\frac{\sigma_{1,2}v_{1,2}}{m_1m_2},\label{eq:alpha}
\end{align}
where $\sigma_{1,2}$ and $v_{1,2}$ are the collisional cross-section
and the relative velocity between the
two grains (we explain how to evaluate $v_{1,2}$ later).
We write the collisional cross-section as
\begin{align}
\sigma_{1,2}=\beta\upi (a_1+a_2)^2,\label{eq:sigma}
\end{align}
where $\beta$ effectively regulates the cross-section relative to the
geometric cross-section. We assume $\beta =1$ for simplicity
(i.e.\ the collision cross-section is determined by the geometric
cross-section), since
in shattering, the grain velocities are so large that the change of
grain collisional cross-section by the Coulomb interaction is negligible.
The second term of the right-hand side in equation (\ref{eq:shat})
indicates the generation of grains with mass $m$ as a result of
collisions between grains with masses $m_1$ and $m_2$.
Note that we count the collision between $m_1$ and $m_2$ twice to
consider the fragments of $m_1$ and $m_2$ separately, and
$\mu_\mathrm{frag}(m;\, m_1,\, m_2)$ describes the mass
distribution of fragments (including the remnant)
as a result of fragmentation of a grain with mass $m_1$ in a collision with 
a grain with mass $m_2$. Because shattering conserves the total dust
mass, the following equation should hold:
\begin{align}
\int_0^\infty\mu_\mathrm{frag}(m;\, m_1,\, m_2)\,
\mathrm{d}m =m_1.
\end{align}
The functional form of $\mu_\mathrm{frag}$ will be given later.
%%where $m_\mathrm{f,min}$ and $m_\mathrm{f,max}$ are the minimum and
%%maximum fragment masses determined later.
To calculate the evolution by shattering, we also need to specify the
grain velocities, which are given in what follows.

The grain velocity is determined by complicated interaction
between the dust grain and the ambient magnetized gas.
\citet{Asano:2013aa} used the data calculated by
\citet{Yan:2004aa}, who provide the grain velocities
for some representative ISM phases.
However, the physical conditions of the ISM in the course of
galaxy evolution are not necessarily represented by those ISM phases.
Thus, it would be convenient if we could find an analytic
(or empirical) formula that correctly reflects some characteristic
dependence on the ambient physical condition and the grain radius
for the purpose of implementation into a hydrodynamic simulation.
For this purpose, we adopt the following formula for the grain velocity as
a function of grain radius (Appendix \ref{app:vel}):
\begin{align}
v_\mathrm{gr}(a) &= 1.1\mathcal{M}^{3/2}\left(
\frac{a}{0.1~\micron}\right)^{1/2}\left(\frac{T_\mathrm{gas}}{10^4~\mathrm{K}}\right)^{1/4}
\left(\frac{n_\mathrm{H}}{1~\mathrm{cm}^{-3}}
\right)^{-1/4}\nonumber\\
&\times \left(\frac{s}{3.5~\mathrm{g~cm}^{-3}}\right)^{1/2}~\mathrm{km~s}^{-1},
\label{eq:vel}
\end{align}
where $\mathcal{M}$ is the Mach number of the largest-eddy velocity
(which is practically used here as an adjusting parameter for the grain
velocity). The functional form correctly catches the general features
of the dependence on various quantities at least qualitatively in the following
senses: larger grains tend to have larger velocities because they tend to be
coupled with larger-scale gas motion; the grain velocities are higher in
a higher-temperature environment because the characteristic velocity
is higher; the grain velocities tend to be higher in a less dense medium
because of weaker gas drag (i.e.\
dust grains are less trapped on small scales on which the velocity dispersion
of gas is small); and
heavier grains tend to have higher velocities because of their larger
inertia (they are more difficult to be stopped by gas drag).
Although the
inclusion of magnetic field makes the dependence complicated,
the above dependences on various physical quantities does not
change qualitatively \citep{Yan:2004aa}.
In the diffuse ISM with $n_\mathrm{H}\sim 0.1$--0.3~cm$^{-3}$
and $T_\mathrm{gas}\sim 6000$--8000 K,
large grains ($a\sim 0.1$--1 $\micron$) achieve a velocity of
$\sim 10$ km s$^{-1}$. Thus, we adopt $\mathcal{M}=3$
for shattering. Here $\mathcal{M}$ effectively includes the
extra grain acceleration beyond the typical thermal velocity through the
interaction with magnetic field.

In considering the collision rate between two grains
with $v_\mathrm{gr}=v_1$ and $v_2$, we estimate the relative velocity $v_{1,2}$ by
\begin{align}
v_{1,2}=\sqrt{v_1^2+v_2^2-2v_1v_2\mu_{1,2}\,}\,,\label{eq:rel_vel}
\end{align}
where $\mu =\cos\theta$ ($\theta$ is an angle between the two grain velocities)
is randomly chosen between $-1$ and 1 in every calculation of $\alpha$
\citep{Hirashita:2013aa}.

For the grain size distribution of the fragments, A13 followed
\citet{Jones:1996aa} \citep[see also][]{Hirashita:2009ab}.
Formation of fragments in their model depends on some material
properties in a somewhat complicated way.
\citet{Hirashita:2013ab} argued that the most important parameter
is the velocity threshold for the catastrophic fragmentation
(defined as the fragmentation in which more than half of the grain is
disrupted). \citet{Kobayashi:2010aa}'s formalism is simply described with only
one material parameter ($Q_\mathrm{D}^\star$ introduced later),
which is related to the velocity threshold for catastrophic fragmentation.
Thus, we adopt \citet{Kobayashi:2010aa}'s formulation for its simplicity,
and summarize it in what follows.

First, we determine the total mass of the fragments by estimating how
much fraction of the dust grain is disrupted in a collision.
Now we consider a collision of two dust grains with masses $m_1$
and $m_2$.
We follow \citet{Kobayashi:2010aa}'s model, which assumes that the
disrupted mass (ejected mass) from $m_1$ is proportional to
\begin{align}
\phi\equiv\frac{E_\mathrm{imp}}{m_1Q_\mathrm{D}^\star},
\end{align}
where
\begin{align}
E_\mathrm{imp}=\frac{1}{2}\frac{m_1m_2}{m_1+m_2}v_\mathrm{1,2}^2,
\end{align}
is the impact energy between $m_1$ and $m_2$, and $Q_\mathrm{D}^\star$
is the specific impact energy that causes the catastrophic disruption
(i.e.\ the disrupted mass is $m_1/2$). Using $\phi$, the ejected mass,
$m_\mathrm{ej}$ is estimated as
\begin{align}
m_\mathrm{ej}=\frac{\phi}{1+\phi}m_1.
\end{align}
This satisfies the following behaviours at two extremes:
$m_\mathrm{ej}\sim E_\mathrm{imp}/Q_\mathrm{D}^\star$ for
$\phi\ll 1$ (weak collision) and
$m_\mathrm{ej}\sim m_1$ for $\phi\gg 1$ (strong collision).
As argued in \citet{Hirashita:2013ab}, we estimate that
$Q_\mathrm{D}^\star\simeq P_1/(2s)$, where $P_1$ is the critical
pressure given by \citet{Jones:1996aa} ($P_1=3\times 10^{11}$
and $4\times 10^{10}$ dyn cm$^{-2}$ for silicate and graphite,
respectively; and we adopt the silicate value in this paper).

Next, we set the grain size distribution of shattered fragments.
We assume a power-law size distribution with an index of
$\alpha_\mathrm{f}$, which means that the index of mass distribution is
$(-\alpha_\mathrm{f}+1)/3$. We adopt $\alpha_\mathrm{f}=3.3$
in this paper \citep{Jones:1996aa}, but we note that the value of
$\alpha_\mathrm{f}$ is not essential in
determining the resulting grain size distribution {as long as
$\alpha_\mathrm{f}<4$} \citep{Hirashita:2013ab}.
The maximum and minimum grain masses of the fragments,
$m_\mathrm{f,max}$ and $m_\mathrm{f,min}$, respectively, are assumed to be
\citep{Guillet:2011aa}
\begin{align}
m_\mathrm{f,max} &= 0.02m_\mathrm{ej},\\
m_\mathrm{f,min} &= 10^{-6}m_\mathrm{f,max}
\end{align}
[or the maximum and minimum fragment radii,
$a_\mathrm{f,max}=(0.02m_\mathrm{ej}/m_1)^{1/3}a_1$
and $a_\mathrm{f,min}=0.01a_\mathrm{f,max}$, respectively].
The minimum size is assumed to
be $\sim 1/100$ times the maximum size, which is roughly consistent with the
treatments in \citet{Jones:1996aa} and \citet{Hirashita:2009ab}.
%%The grains whose masses are below $m_\mathrm{min}$ are removed from
%%the calculation (this removal also does not affect the results significantly).
In the end, we obtain the fragment mass distribution including the remnant
of mass $m_1-m_\mathrm{ej}$ as
%%$\mu_\mathrm{frag}(m;\, m_\mathrm{ej})\propto m^{(-\alpha_\mathrm{f}+1)/3}$.
%%Thus, we obtain, including the remnant
\begin{align}
\mu_\mathrm{frag}(m,\, m_1,\, m_2) &=
\frac{(4-\alpha_\mathrm{f})m_\mathrm{ej}m^{(-\alpha_\mathrm{f}+1)/3}}{3\left[
m_\mathrm{f,max}^\frac{4-\alpha_\mathrm{f}}{3}-
m_\mathrm{f,min}^\frac{4-\alpha_\mathrm{f}}{3}\right]}\nonumber\\
&+ (m_1-m_\mathrm{ej})\delta (m-m_1+m_\mathrm{ej}),\label{eq:frag}
\end{align}
where $\delta (\cdot )$ is Dirac's delta function.
In reality, we use the discrete formalism for the distribution of fragments
(Appendix \ref{app:shattering}).
Grains which become smaller than the minimum grain size (radius $a_\mathrm{l}$)
are removed.
%%If $m_\mathrm{f,max}$ calculated above is smaller than $m_\mathrm{min}$,
%%we remove all the fragments
%%(however, the removed fraction is so small
%%that it does not affect the results significantly). 

\subsection{Coagulation}\label{subsec:coagulation}

Coagulation occurs in the dense ISM ($n_\mathrm{H}\gtrsim 100$ cm$^{-3}$),
where grain velocities induced by
turbulence are small enough to allow the grains to stick with each other
in grain--grain collisions. The evolution of grain size distribution by coagulation
is written in a similar way to that by shattering as
\begin{align}
\left[\frac{\partial\rho_\mathrm{d}(m,\, t)}{\partial t}\right]_\mathrm{coag}
= -m\rho_\mathrm{d}(m,\, t)\int_0^\infty\alpha (m_1,\, m)\rho_\mathrm{d}(m_1,\, t)
\mathrm{d}m_1\nonumber\\
+ \int_0^\infty\int_0^\infty\alpha (m_1,\, m_2)\rho_\mathrm{d}(m_1,\, t)\rho_\mathrm{d}(m_2,\, t)
m_1\delta (m-m_1-m_2)\mathrm{d}m_1\mathrm{d}m_2.\label{eq:coag}
\end{align}
We note that we count the same coagulation twice by
treating two coagulated grains, $m_1$ and $m_2$, separately
in the second term on
the right-hand side.\footnote{In
equation (\ref{eq:coag}), because of the $\delta$ function,
the integration is executed under constraint $m_2=m-m_1$.
We formally adopt $\rho_\mathrm{d}(m_2,\, t)=0$ when $m_2<0$ (i.e.\ $m_1>m$).
Because the second term on the right-hand side in equation (\ref{eq:coag})
should be symmetric by the interchange of $m_1$ and $m_2$,
we can find that it is also expressed as $\frac{m}{2}\int_0^\infty\alpha (m_1,\, m-m_1)
\rho_\mathrm{d}(m_1,\, t)\rho_\mathrm{d}(m-m_1,\, t)\mathrm{d}m_1$.}
For $\alpha$ and $\sigma$, we use the same expressions as in
equations (\ref{eq:alpha}) and (\ref{eq:sigma}), respectively.
For coagulation, $\beta$ in equation (\ref{eq:sigma}) includes the sticking
coefficient (we adopt $\beta =1$ also for coagulation).
We neglect the change of grain cross-section due to the Coulomb
interaction, since the kinetic energy of grains is much higher than
the typical Coulomb potential energy.\footnote{Detailed grain charging
processes could be important for coagulation if the grain motion is Brownian
(thermal) \citep[e.g.][]{Ivlev:2015aa}.
Under the density range of interest ($n_\mathrm{H}\la 10^3$ cm$^{-3}$),
turbulence is unlikely to decay completely \citep{Larson:1981aa}.
Since the grain velocities induced
by turbulence are much higher than the thermal velocities,
the Coulomb potential is not important for coagulation.}

We adopt equation (\ref{eq:vel}) for the grain velocity
(with $\mathcal{M}=1$, which roughly mimics the velocity level calculated
by \citealt{Yan:2004aa} for the dense clouds) and use the
same method to estimate the relative velocity as applied for shattering
(equation \ref{eq:rel_vel}).
%%Since we only consider coagulation in dense and cold gas,
%%the velocities are much smaller than those used for shattering.
A13's model assumed a threshold velocity beyond which
coagulation is prohibited \citep[see also][]{Chokshi:1993aa,Dominik:1997aa}.
As shown in \citet{Asano:2014aa}, however, the Milky Way extinction curve
cannot be explained if coagulation is stopped by such a threshold.
\citet{Hirashita:2014ab} indeed showed that the variation of extinction curves
in the Milky Way is better explained without coagulation threshold.
Therefore,
we assume that coagulation always occurs if two grains meet
in the dense ISM.

\section{Application to a one-zone model}\label{sec:application}

There are some simplifications and modifications in our formulation
compared with A13. Although our main purpose is
to develop a model to be implemented in a hydrodynamical simulation,
it is useful to check if our model produces similar grain size distributions
to those found in previous models (especially A13). Our dust model requires metallicity
evolution (or chemical evolution) as input. A13 developed
an elaborated model for chemical enrichment by adopting detailed
metal yield tables. They also used theoretical calculation results of dust
condensation efficiency for AGB stars and SNe in the literature.
However, as mentioned above,
our formulation avoids adopting a specific dust yield table because of
the uncertainty in such calculations. Therefore, it is not possible to
compare our calculation with A13's under the exactly same condition.
Nevertheless, since the simplification we apply should
not change the model essence, our results should be similar
to theirs.

We simply
mimic A13's chemical enrichment model by using the following
fitting formula to their results:
$Z=0.6({t}/\tau_\mathrm{SF})~\mathrm{Z}_{\sun}$,
where $\tau_\mathrm{SF}$ is the star formation time-scale
($\tau_\mathrm{SF}\equiv M_\mathrm{gas}/\psi$, where $M_\mathrm{gas}$ the
total gas mass and $\psi$ is the star formation rate), which is assumed to be
constant.
This formula comes from their figure 6, which indeed shows that
the metallicity evolves roughly proportionally to the age under
a constant star formation time-scale. Accordingly, we assume that the metals are
supplied at a constant rate as
$\dot{\rho}_Z=\rho_\mathrm{gas}\dot{Z}$.
We adopt $\tau_\mathrm{SF}=5$~Gyr (A13's fiducial value).
Note that the metallicity evolution formula in this paragraph is
just used for the one-zone model adopted in this section.
We solve equation (\ref{eq:basic}) but neglect the last term
(i.e.\ $\dot{\rho}_\mathrm{gas}=0$).

Following A13's model, we separate the ISM into two phases:
the cold and warm phases.
We adopt $T_\mathrm{gas}=10^2$~K and $n_\mathrm{H}=30$~cm$^{-3}$
in the cold medium, and $T_\mathrm{gas}=10^4$~K
and $n_\mathrm{H}=0.3$~cm$^{-3}$ in the warm medium.\footnote{A13 adopted
$T_\mathrm{gas}=6000$ K, but we adopt this value for the pressure equilibrium with the
cold medium. The difference, however, does not affect the results significantly.}
The mass fraction of the cold
phase is denoted as $\eta_\mathrm{cold}$. Accordingly, the
mass fraction of the warm phase is $1-\eta_\mathrm{cold}$.
We consider coagulation and accretion only in the cold phase,
while we calculate shattering only in the warm phase.
Because the model used in this section treats a galaxy as a single-zone object,
the galaxy structure cannot be taken into account. Thus, in a single time-step
$\Delta t$, we calculate coagulation and accretion in
$\eta_\mathrm{cold}\Delta t$, while we compute shattering in
$(1-\eta_\mathrm{cold})\Delta t$.
We consider the other processes (stellar dust production and
dust destruction by SN shocks) in the entire time-step.
%%For the dust condensation efficiency, we adopt $f_\mathrm{in}=0.1$
%%\citep{Kuo:2013aa}.

For SN destruction, we assume that SNe (with progenitor mass
$>8$ M~$_{\sun}$) occur instantaneously
after star formation (i.e.\ the lifetime of SN progenitor is zero).
In this case, the SN rate $\gamma$ is
proportional to the star formation rate ($\psi$). Thus, for the
ratio $M_\mathrm{gas}/\gamma$ appearing in equation (\ref{eq:tau_dest}),
we obtain
$M_\mathrm{gas}/\gamma =M_\mathrm{gas}/(\nu_\mathrm{SN}\psi )
=\tau_\mathrm{SF}/\nu_\mathrm{SN}$, where $\nu_\mathrm{SN}$ is the
proportionality constant between $\psi$ and $\gamma$.
We adopt $\nu_\mathrm{SN}=1.0\times 10^{-2}$ based on the Chabrier
\citep{Chabrier:2003aa} initial mass function
(i.e.\ we adopt $M_\mathrm{gas}/\gamma =5\times 10^7$~yr in this section).

\begin{figure}
 \includegraphics[width=\columnwidth]{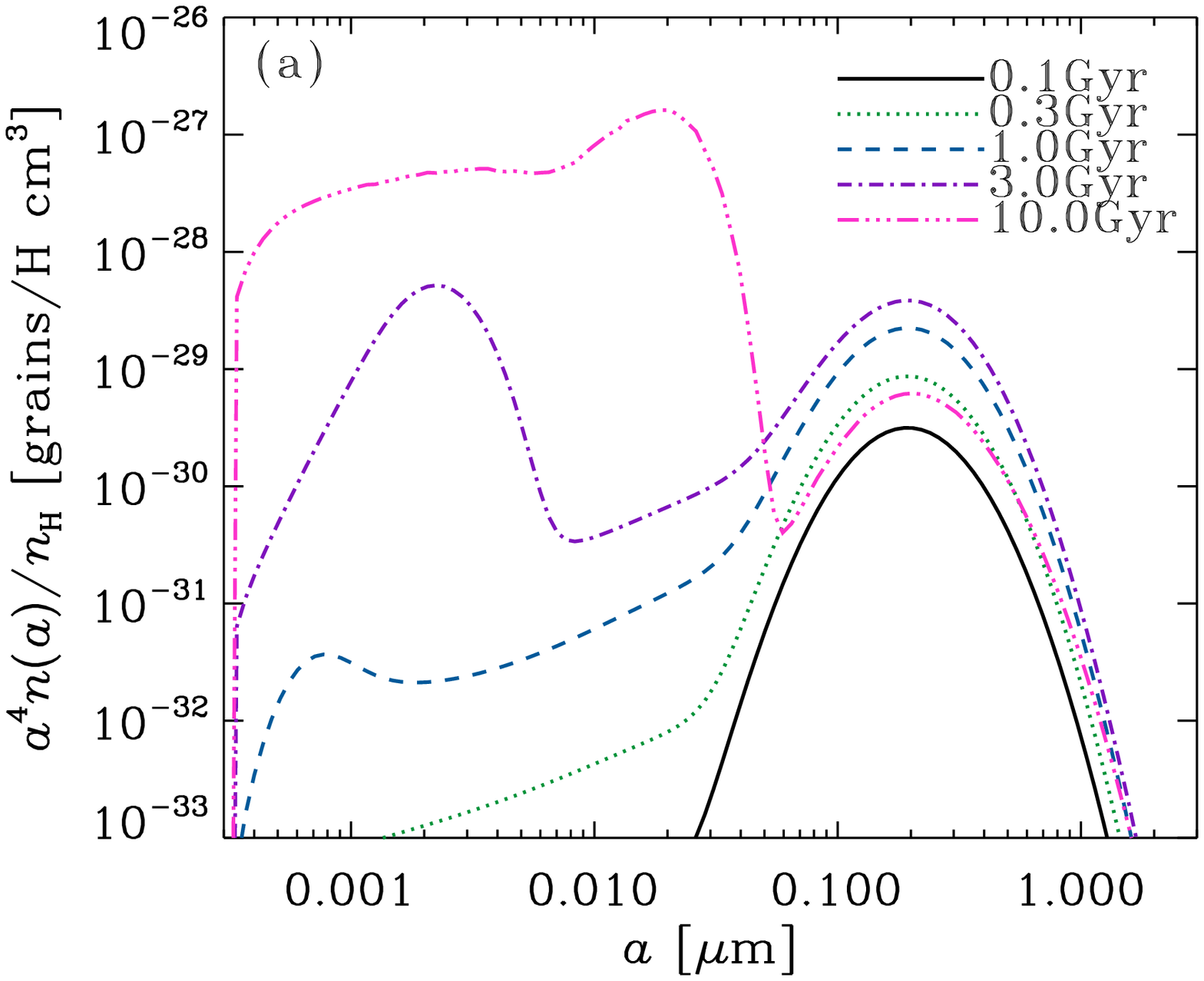}
 \includegraphics[width=\columnwidth]{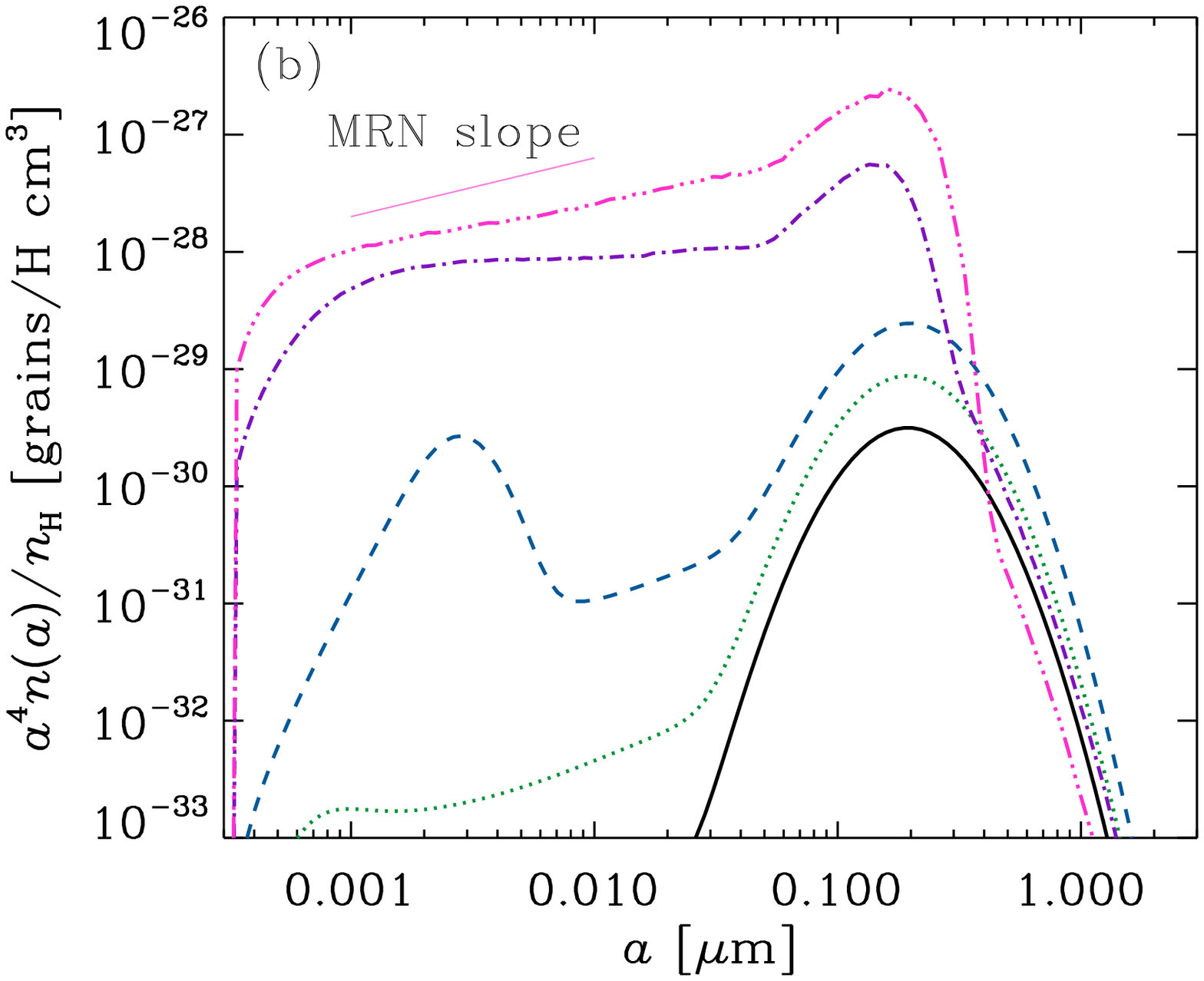}
 \caption{Evolution of grain size distribution for the one-zone model in
 Section \ref{sec:application}.
 The solid, dotted, dashed, dot--dashed, and triple-dot--dashed lines represent
 the results at $t=0.1$, 0.3, 1, 3, and 10 Gyr, respectively. For the vertical
 axis, we present the grain size distribution per hydrogen
 multiplied by $a^4$: the resulting quantity is proportional to the grain mass
 distribution per $\log a$. Thus, if the value of $a^4n(a)$ is high at a certain grain radius,
 it means that the mass is dominated at that grain size.
 Panels (a) and (b) show the standard one-zone model and the dense one-zone model,
 respectively.
 The thin line marked with `MRN slope' in Panel (b) shows the power-law slope of the MRN grain
 size distribution [$n(a)\propto a^{-3.5}$].}
 \label{fig:onezone}
\end{figure}

In Fig.\ \ref{fig:onezone}a, we show the time evolution of grain size distribution.
This model is referred to as the standard one-zone model.
We reproduce A13's evolutionary behaviour of grain size distribution as follows.
%%although there are some slight difference as described below because of our simplification.
The grain size distribution
is dominated by stellar dust production at $t\lesssim 0.3$~Gyr; as a consequence,
the grain size distribution is dominated by large ($a\sim 0.2~\micron$) grains.
At $t\gtrsim 0.3$Gyr, shattering begins to produce small grains. After
$t\sim 1$ Gyr, dust growth by
accretion drastically increases the small grain abundance. As noted by A13,
accretion has a prominent effect on small grains, which have shorter
accretion time-scales than large grains \citep[equation \ref{eq:tau_acc}; see also][]{Kuo:2012aa}.
At $t\sim 3$--10 Gyr, accretion continues to increase the small grain
abundance, while coagulation pushes the grains to larger sizes.
Because the increased abundance of small grains enhances shattering of
large grains, the abundance of the largest grains decreases at $t\sim 3$--10 Gyr.

There are some slight differences from A13's results.
At young ages ($t\la 0.3$ Gyr), the peak of $a^4n(a)$ lies around
$\sim 0.5~\micron$ in A13. This is due to the difference in the adopted size
distribution of dust grains produced by stars, especially by SNe. However, the size
distribution of dust grains formed by SNe is uncertain in the sense that it
is affected by the ambient ISM density and grain species \citep{Nozawa:2007aa}.
At old ages ($t\sim 10$~Gyr), there is a prominent
peak of $a^4n(a)$ at $a\sim 0.01~\micron$ in A13's result, while our model
predicts a rather flat shape of $a^4n(a)$ between $a\sim 0.001$ and $0.02~\micron$.
This is because we do not impose any coagulation threshold as
explained in Section \ref{subsec:coagulation}: in A13's
model, coagulation is stopped around $\sim 0.01~\micron$ because
larger grains have higher velocity than the threshold.
Since we do not impose the threshold, coagulation occurs further in our model.
However, coagulation is not strong enough to produce 0.1 $\micron$
grains even in our model.
%%because the density of the dense medium is relatively low (30 cm$^{-3}$).
There is a dip around $a\sim 0.1~\micron$ caused by the size gap
between the coagulated grains and the stardusts, which is also seen
in A13. In A13, there is another dip created by shattering at sub-micron radii, but this is
simply caused by their adopted grain velocities, which have a
steep increase at sub-micron sizes. This steep dependence of grain
velocity on grain radius
should be smoothed if we consider a variety of gas densities.
As a consequence, we expect that the dip at sub-micron radii in A13 would be
eliminated in reality, and that the grain size distribution approaches what
we predict in this paper.
%%In summary, the different behaviour is caused by the specific model
%%features; therefore, we conclude that, within the current
%%knowledge of dust evolution, our model produces consistent evolution
%%of grain size distribution to that calculated by A13.

As pointed out by \citet{Nozawa:2015aa}, inclusion of a denser medium
is necessary to produce the MRN-like grain size distribution that reproduces
the Milky Way extinction curve. In their scenario, creation of grains with
$a\sim 0.1~\micron$ by coagulation is essential.
Therefore, we examine another model in which we assume the gas density
and temperature
in the dense phase to be $n_\mathrm{H}=300$~cm$^{-3}$ and
$T_\mathrm{gas}=25$ K \citep{Nozawa:2015aa}.\footnote{Precisely speaking,
\citet{Nozawa:2015aa} included such a dense gas phase in addition to the
above two phases. Since it is not convenient to add such a component in our
code, we simplified the formulation by neglecting the component with
$n_\mathrm{H}=30$ cm$^{-3}$. This simplified treatment is sufficient for the
purpose of this section.}
This model is referred to as the dense one-zone model.
In Fig.\ \ref{fig:onezone}b, we show the evolution of grain size
distribution for the dense one-zone model.
Compared with the standard one-zone model, the dense one-zone model
shows quicker dust growth. At $t=0.3$ Gyr, the two models have little
difference because stellar dust production and shattering are not affected
by the dense medium. At $t=1$ Gyr, the bump at $a\sim 0.003~\micron$
produced by accretion is higher in the dense one-zone model because the denser
condition is suitable for efficient accretion.
At $t=3$--10 Gyr, the bump created by accretion is smoothed by coagulation,
which produces grains even larger than $a\sim 0.1~\micron$.
At $t=10$ Gyr, the slope of the grain size distribution is similar to that
of the MRN distribution. Interestingly, the upper cut-off of grain radius is
also consistent with the value assumed in the MRN grain size distribution
($\sim 0.25~\micron$). Therefore, our model also
confirms the conclusion by \citet{Nozawa:2015aa}, who pointed out
the necessity of the dense medium (i.e.\ efficient accretion and coagulation)
in reproducing the MRN grain size distribution.

\section{Application to a galaxy simulation}\label{sec:result}

\subsection{Simulation}\label{subsec:simulation}

Our development of grain size evolution
model is aimed at implementation in galaxy simulations.
For the first step, we here post-process the simulation of an isolated
spiral galaxy in A17. We briefly review their
simulation, and refer the interested reader to A17 for details.

We use the modified version of \textsc{gadget3} $N$-body/SPH code
(the original version was introduced by \citealt{Springel:2005aa}, and
the modified version is referred to as \textsc{gadget3-osaka}), and
our simulation includes dark matter, gas and star particles. 
We install the Grackle\footnote{https://grackle.readthedocs.org/} chemistry
and cooling library \citep{Bryan:2014aa,Kim:2014aa,Smith:2017aa}
to solve non-equilibrium primordial chemistry network
{for H, D, He, H$_2$ and HD.
This allows us to compute the gas properties down to low temperatures
$\sim 100$ K and up to higher densities $n\sim 100$ cm$^{-3}$.}
We adopt the
same initial condition as used in
the low-resolution model of \texttt{AGORA} simulations \citep{Kim:2014aa,Kim:2016aa}
{(see also table 1 of A17).
To `construct' a disc galaxy, they included three components in the
initial condition: halo, stellar
disc, gas disc and bulge with the total mass of each component being
$1.25\times 10^{12}$, $4.30\times 10^9$, $8.59\times 10^9$, and
$3.44\times 10^{10}$ M$_{\sun}$, respectively. The numbers of particles
are $10^5$ for the halo, stellar disc, and gas disc, and $1.25\times 10^4$ for the
bulge.}
We apply the minimum gravitational softening length $\epsilon_\mathrm{grav}=80$ pc
and we allow the baryons to collapse to 10 per cent of this value.
{
We find that the variable gas smoothing length reaches a minimum value of $\sim 22$ pc
with our models of gas cooling, star formation and feedback.}
The star formation is assumed to occur in a local free-fall time with an efficiency of 0.01,
{and the star particles are stochastically created from gas particles as described
in \citet{Springel:2003aa}, consistently with the SFR density.}
A17 only considered metal production and stellar feedback by Type II SNe
{according to \citet{Kim:2014aa} and \citet{Todoroki:2014aa}.}
Following \citet{Aoyama:2018aa}, we newly include the metal and dust production by
Type Ia SNe and AGB stars in addition to Type II SNe. The formation of
various metal elements is treated by implementing the \texttt{CELib} package
\citep{Saitoh:2017aa},
{and the metals are injected according to the lifetimes of
the progenitors and the metal yields (see the references in \citealt{Saitoh:2017aa}).
At the same time of metal production, stellar feedback is also considered
by depositing progenitor-dependent energy given by \texttt{CELib} in the
neighbouring gas particles.
The fractions of distributed energy and metal mass among the particles are determined by
the weight proportional to the SPH kernel in the same way as A17.}
This new implementation based on \texttt{CELib} does not cause a significant difference from
A17 as far as this paper is concerned.

\subsection{Post-processing for dust evolution}\label{subsec:post-process}

\begin{figure}
 \includegraphics[width=0.98\columnwidth]{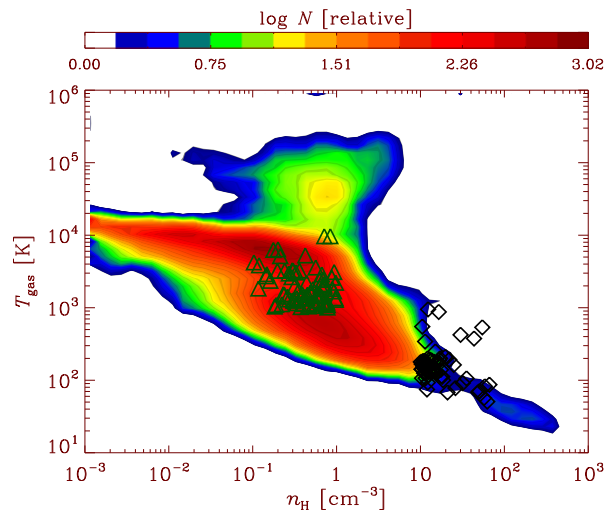}
 \caption{Distribution of the sampled gas particles at $t=10$ Gyr
 on the phase ($T_\mathrm{gas}$--$n_\mathrm{H}$) diagram.
 The green triangles and black diamonds show the
 75 cold/dense and 77 warm/diffuse gas particles
 (dense and diffuse sample), respectively. The colour
 map shows the distribution of all the gas particles in the simulation at
 $t=10$ Gyr. The colours correspond to the logarithmic number density
 of the gas particles as shown in the colour bar. Some data points located
 in the white region are isolated data points (the coloured region shows
 the area where data points are concentrated).}
 \label{fig:sample}
\end{figure}

We sample some SPH gas particles (referred to as
gas particles) and investigate the dust evolution on those particles
in detail.
This implicitly assumes that the dust is dynamically coupled with the
gas through gas drag.
%%We choose 30 gas particles from each of the cold/dense
%%and warm/diffuse ISM in the snapshot at $t=10$ Gyr.
We choose particles in the snapshot at $t=10$~Gyr and trace back
the history of those particles.
The cold/dense (warm/diffuse) ISM is defined as
$n_\mathrm{H}>10$~cm$^{-3}$ and $T_\mathrm{gas}<10^3$~K
($0.1<n_\mathrm{H}<1$ cm$^{-3}$ and $10^3<T_\mathrm{gas}<10^4$ K), and
extract {75 (77) particles for the cold/dense (warm/diffuse)
phases in the following way.}
We ignore the particles at $R<0.1$ kpc ($R$ is the radius
from the galaxy centre in {the projection onto} the disc plane), where we find a concentration
of gas particles whose physical condition is not typical of the galactic disc
($n_\mathrm{H}>100$ cm$^{-3}$ and $Z\gtrsim 3$ Z$_{\sun}$)
{but could be typical of the `galactic centre'}.
Thus, we choose gas particles located at $R=0.1$--4 kpc
{to examine the dust evolution in a galactic disc}.
The outer radius, $R=4$ kpc, is determined by the fact that we rarely
find cold/dense gas particles at $R>4$ kpc.
To extract the gas contained in the disc, we also restrict the
height as $-0.3<z<0.3$ kpc ($z$ is the height from the galactic disc plane).
To {avoid selecting particles which experienced peculiar chemical
enrichment not typical of the galactic disc (e.g.\ halo-origin,
galactic-centre-origin, etc.),} we only select gas particles with
$0.5<Z<2$ Z$_{\sun}$ in the 10 Gyr snapshot (however, only a small
number of gas particles are excluded by this metallicity criterion).
Among the particles satisfying
these criteria, we randomly extracted 75 and 77 particles from the cold/dense and
warm/diffuse gas particles.
These numbers are determined to obtain statistically robust
conclusions. Direct implementation of our model to the simulation (i.e.\
all the gas particles) is left for future work.
The extracted gas particles are shown on the phase
($T_\mathrm{gas}$--$n_\mathrm{H}$) diagram in Fig.\ \ref{fig:sample}.
The 75 cold/dense and 77 warm/diffuse particles are referred to as the
dense and diffuse sample, respectively.

In order to calculate the evolution of grain size distribution, we
record the time evolution of
gas density ($n_\mathrm{H}$), gas temperature ($T_\mathrm{gas}$)
and metallicity ($Z$) for each gas particle.
In addition, we also need
the timing and number of SNe sweeping the gas particle for the
purpose of estimating dust destruction by SNe (see below).
Since the number of SNe is a discrete integer number, we count it up
($N_\mathrm{SN}$ is the cumulative number of SNe affecting the
gas particle of interest; see Section 2.2 of A17 for how to estimate the
number of SNe on each gas particle).

Using the metallicity evolution of each gas particle, we
calculate $\dot{\rho}_Z=\mu m_\mathrm{gas}n_\mathrm{gas}\dot{Z}$,
which is used to estimate the stellar dust production through
equation (\ref{eq:stellar}).
%%For the dust condensation efficiency, we adopt $f_\mathrm{in}=0.1$
%%\citep{Kuo:2013aa}.
For dust destruction by SNe, we
evaluate the SN rate by $\gamma =\Delta N_\mathrm{SN}/\Delta t$,
where $\Delta N_\mathrm{SN}$ is the number of SNe hitting on the gas
particle in the time
interval $\Delta t$ [$\Delta t$ is the time-step, and
$\Delta N_\mathrm{SN}=N_\mathrm{SN}(t+\Delta t)-N_\mathrm{N}(t)$].
We use the mass of
the gas particle for $M_\mathrm{gas}$ in equation (\ref{eq:tau_dest}).

It is generally difficult to spatially resolve dense clouds where grain growth
by accretion and coagulation occurs. Thus, we adopt a subgrid model
for coagulation and accretion following A17.
We `turn on' accretion and coagulation for gas particles
that satisfy $n_\mathrm{H}>10$~cm$^{-3}$ and $T_\mathrm{gas}<1000$ K.
%%We refer to those gas particles dust-growing particles.
We assume that the mass fraction of $f_\mathrm{dense}$
of such a dense particle is condensed into dense clouds
with $n_\mathrm{H}=n_\mathrm{H,dense}=10^3$ cm$^{-3}$ and
$T_\mathrm{gas}=T_\mathrm{gas,dense}=50$ K on
subgrid scales and that accretion and coagulation occur only in those dense clouds.

Since shattering preferentially occurs in the diffuse ISM
\citep{Yan:2004aa}, we turn on shattering only in gas particles
with $n_\mathrm{H}<1$ cm$^{-3}$.
No subgrid treatment is necessary for shattering, since the diffuse ISM
is spatially resolved.

\subsection{Analysis of grain size evolution}\label{subsec:SPHgrainsize}

\begin{figure*}
 \includegraphics[width=1\columnwidth]{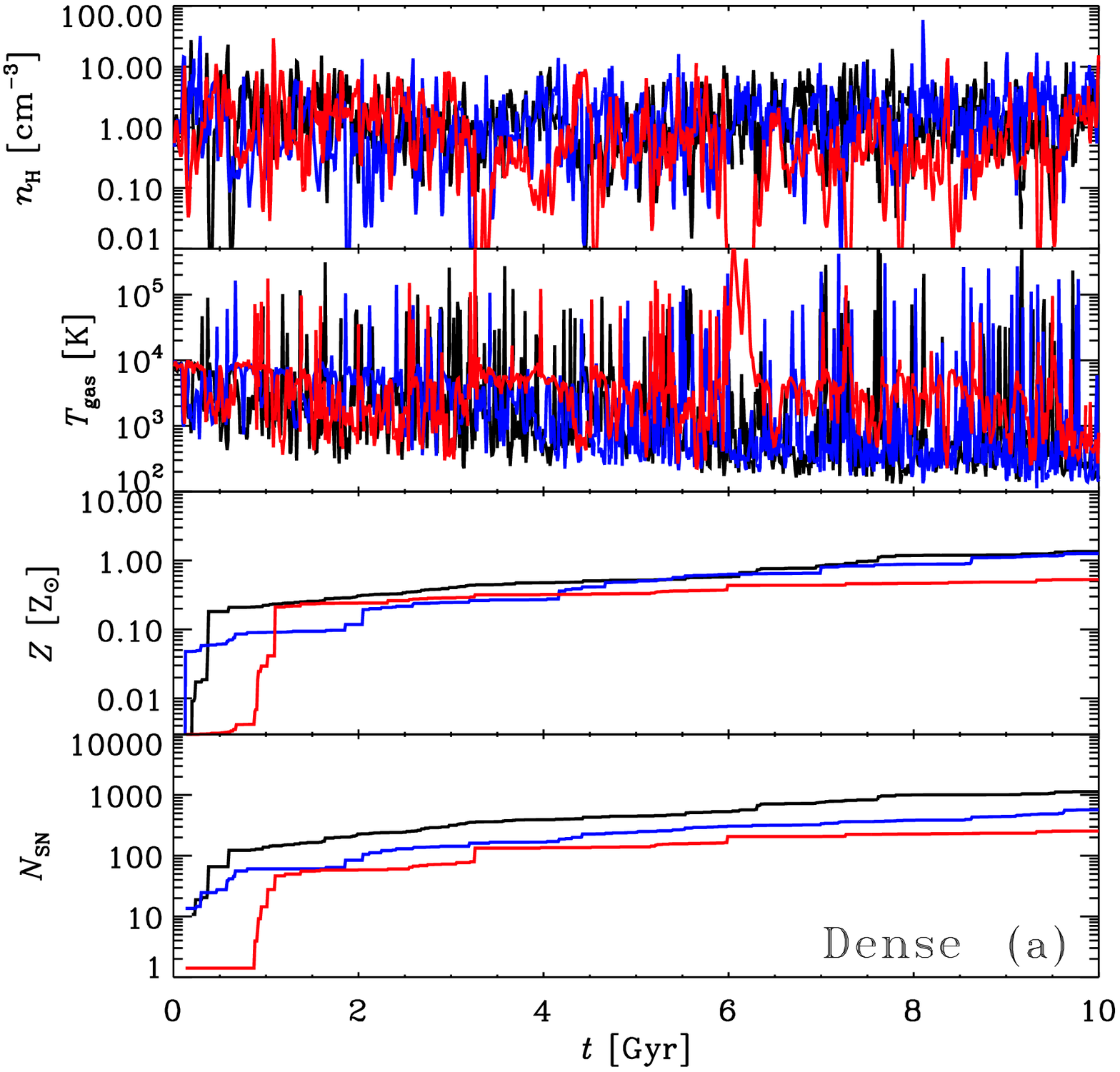}
 \includegraphics[width=1\columnwidth]{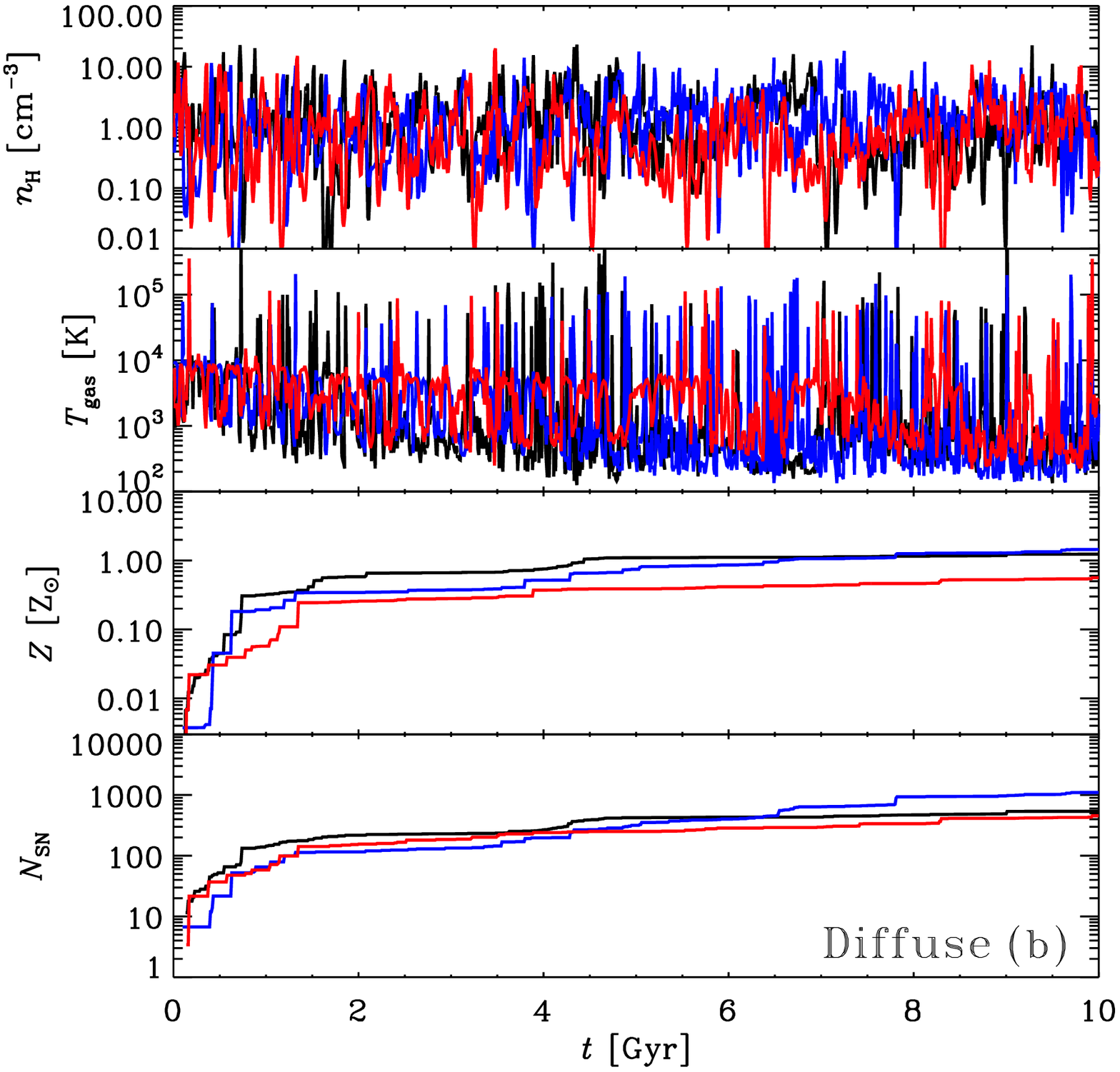}
 %% black blue red: 75148 50583 35299 for dense
 %% 42043 22316 70931 for diffuse
 \caption{Time evolution of physical quantities in {three randomly chosen} gas
 particles in (a) the dense sample and (b) the diffuse sample. The different
 colours (black, red, and blue) show different gas particles.
 The quantities shown from top to bottom are the hydrogen number density,
 gas temperature, metallicity, and cumulative number of SNe.}
 \label{fig:history}
\end{figure*}

\begin{figure*}
 \includegraphics[width=0.66\columnwidth]{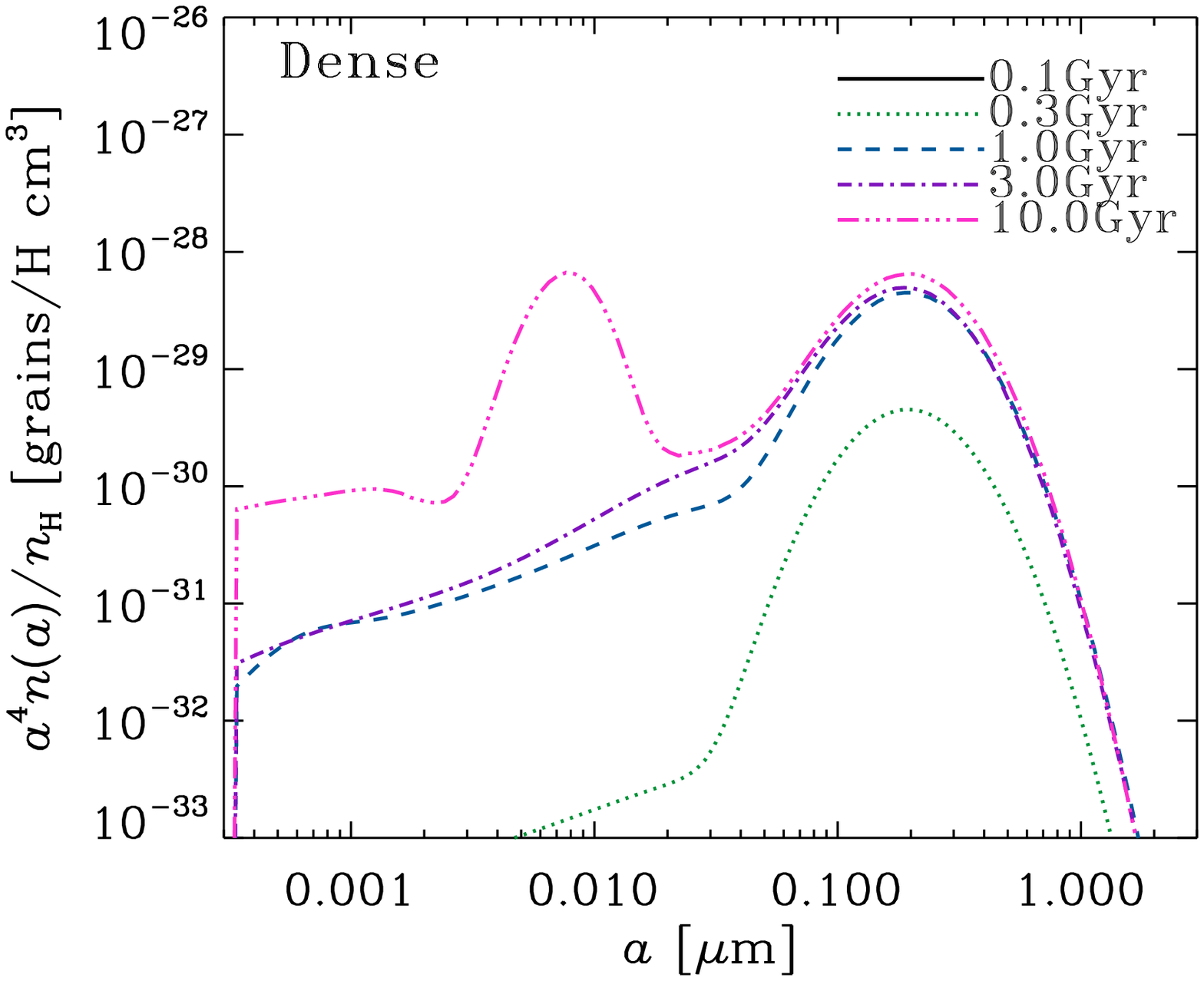}
 \includegraphics[width=0.66\columnwidth]{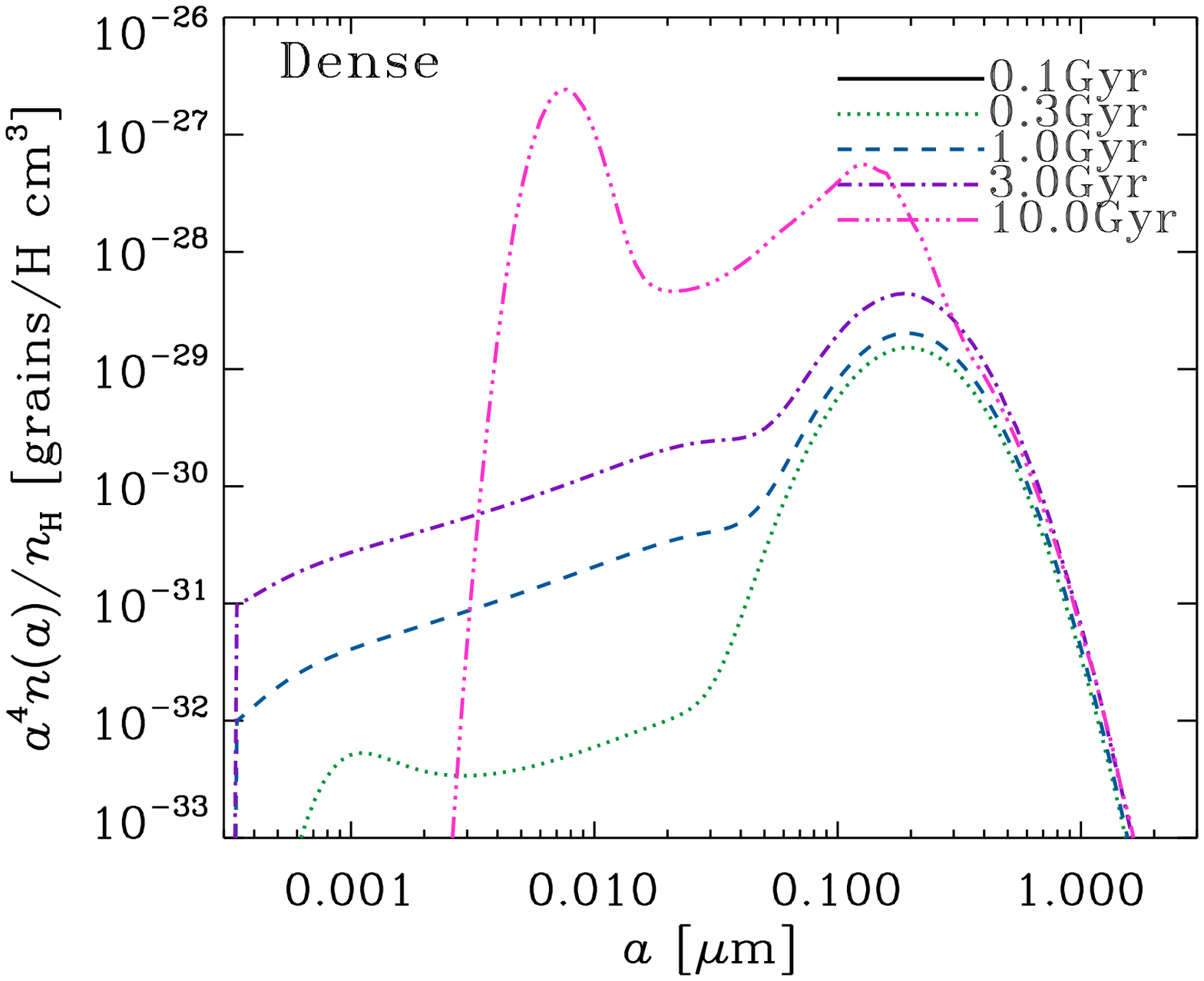}
 \includegraphics[width=0.66\columnwidth]{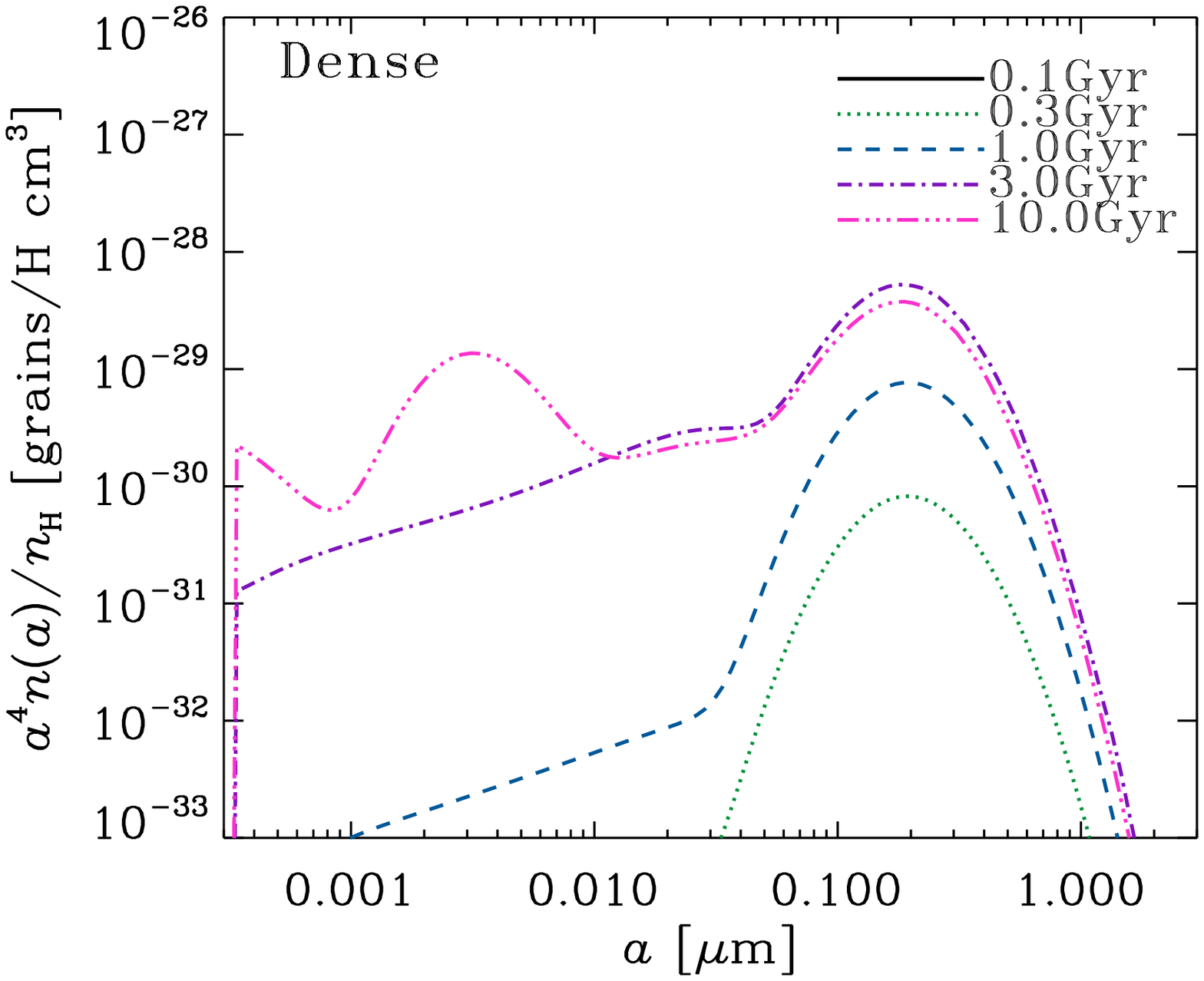}
 \caption{Grain size distributions at $t=0.1$, {0.3}, 1, 3, and 10 Gyr
 for the solid, dotted, dashed, dot--dashed, and triple-dot--dashed lines,
 respectively. The left, middle, and right panels show the evolution of
 the dense gas particles shown in black, blue, and red in Fig.\ \ref{fig:history}a,
 respectively.}
 \label{fig:size_ev_dense}
\end{figure*}

\begin{figure*}
 \includegraphics[width=0.66\columnwidth]{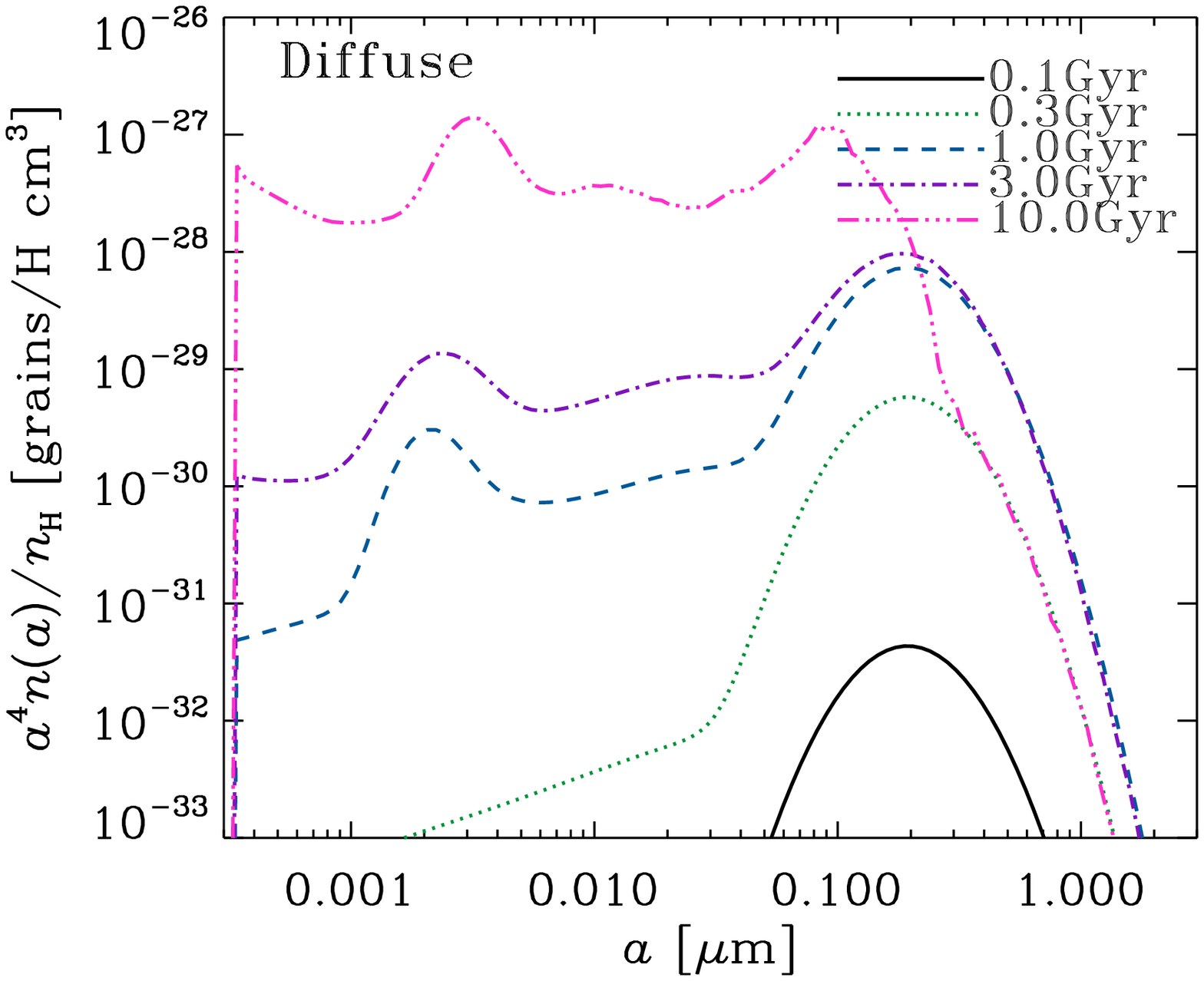}
 \includegraphics[width=0.66\columnwidth]{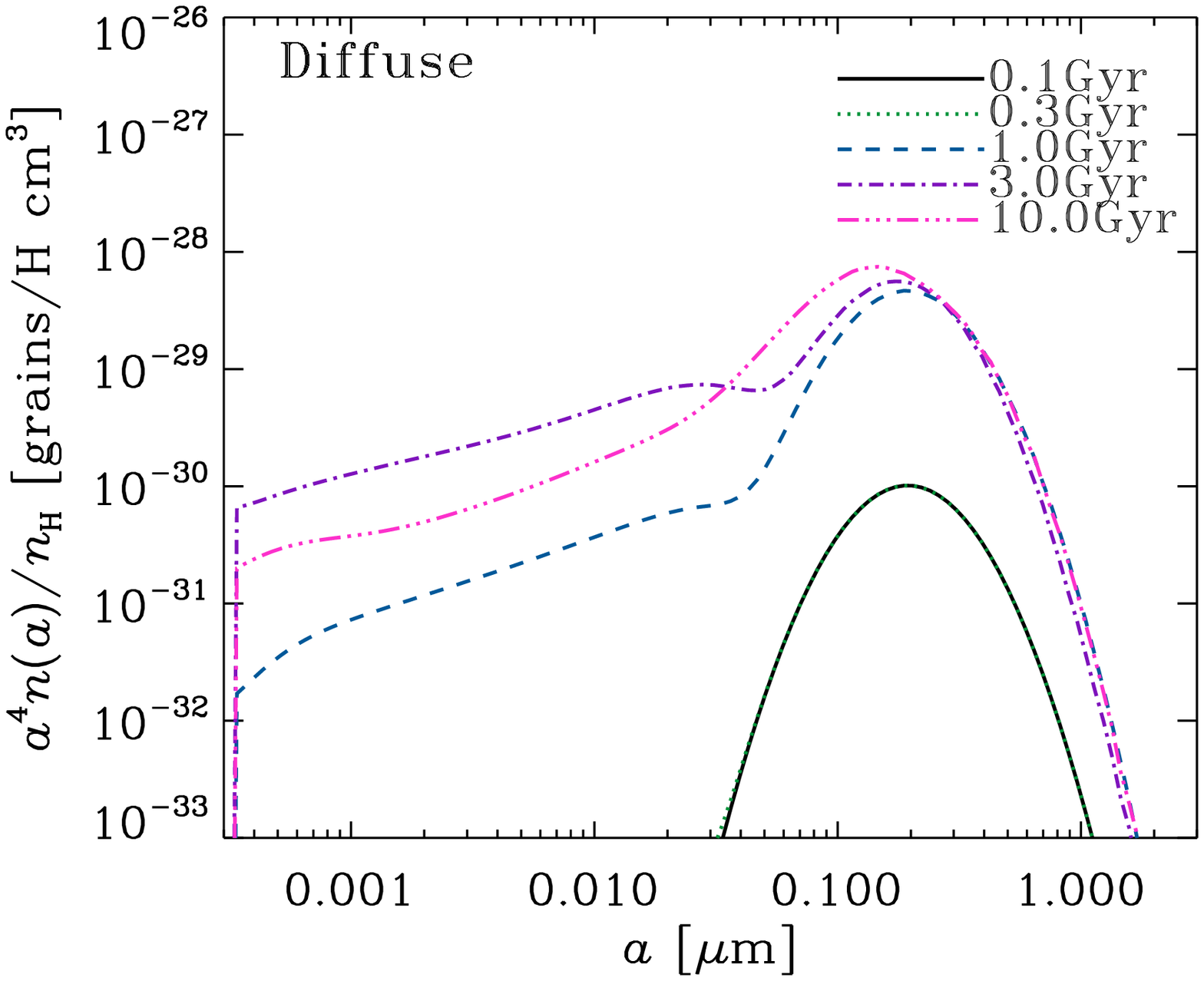}
 \includegraphics[width=0.66\columnwidth]{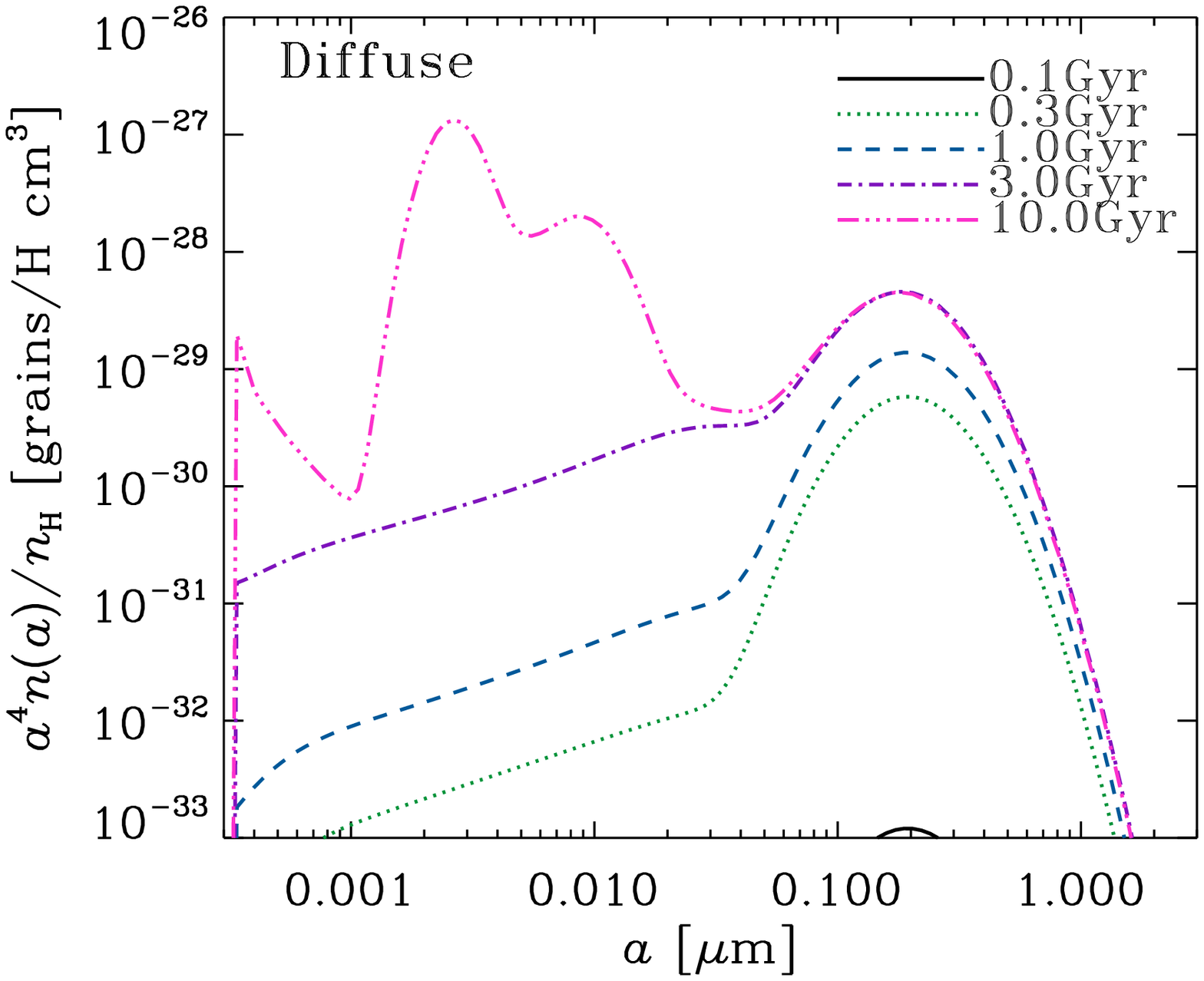}
 \caption{Same as Fig.\ \ref{fig:size_ev_dense} but for the diffuse gas particles
 shown in Fig.\ \ref{fig:history}b.}
 \label{fig:size_ev_diffuse}
\end{figure*}

In order to show the history of the physical conditions which
each gas particle experiences,
we present the evolution of the relevant quantities for
{three randomly chosen} gas particles in the dense and diffuse sample
(recall that they are sampled according to the physical state at $t=10$ Gyr)
in Fig.\ \ref{fig:history}. We find that the gas density and temperature
fluctuate significantly regardless of the physical state at $t=10$ Gyr.
There is no systematic difference in the evolution of metallicity and
SN number between the two samples. The metallicity increase
correlates with the number of SNe because SNe supply metals.

We show the evolution of grain size distribution for the above three chosen
gas particles
in the dense sample in Fig.\ \ref{fig:size_ev_dense}. We observe
an evolutionary trend common for all the three particles and similar to
the behaviour found for the one-zone models in Section \ref{sec:application}:
in the early phase, the dust is dominated by large grains produced by stars.  At
$t\sim 0.3$--1 Gyr, a tail develops  toward the small grain size by
shattering. At the later stage, accretion and coagulation drastically affect the
grain size distribution: accretion increases the small grain abundance and creates a
bump as noted in Section \ref{sec:application}, while coagulation
tends to deplete small grains and increase large grains.
Although these evolutionary tendencies are common for all the three
particles, there is a large variety in the grain size distribution, especially in the
later phase, mainly because dust growth processes (accretion and coagulation)
that dramatically impact the grain size distribution are sensitive to the
history (i.e.\ when and how long the dust is included in the cold/dense phase).

In Fig.\ \ref{fig:size_ev_diffuse}, we show the evolution of grain size distribution
for the above three chosen particles in the diffuse sample. We broadly find
an evolutionary trend similar to that
found for the one-zone models in Section \ref{sec:application}
and for the dense sample above, although the variety is large among the gas
particles.
%%that is, the dominance of large grains in the early epoch and the buildup of
%%small grains by shattering and accretion in the later phase.
%%We observe a large variety in the grain size distribution, especially at the
%%later phase, and there seems to be no systematic difference from the
%%dense sample (Fig.\ \ref{fig:size_ev_dense}).
Statistical comparison between
the diffuse and dense samples is presented later.
Interestingly, the second case in Fig.\ \ref{fig:size_ev_diffuse} shows that
small grains are more depleted at 10 Gyr than at 3 Gyr. This is because of
the destruction by SNe in the last few Gyr; indeed, this case corresponding to
the blue line in Fig.\ \ref{fig:history}b shows a larger increase in $N_\mathrm{SN}$
at $t=7$--10 Gyr compared with the other cases.

\begin{figure*}
 \includegraphics[width=1\columnwidth]{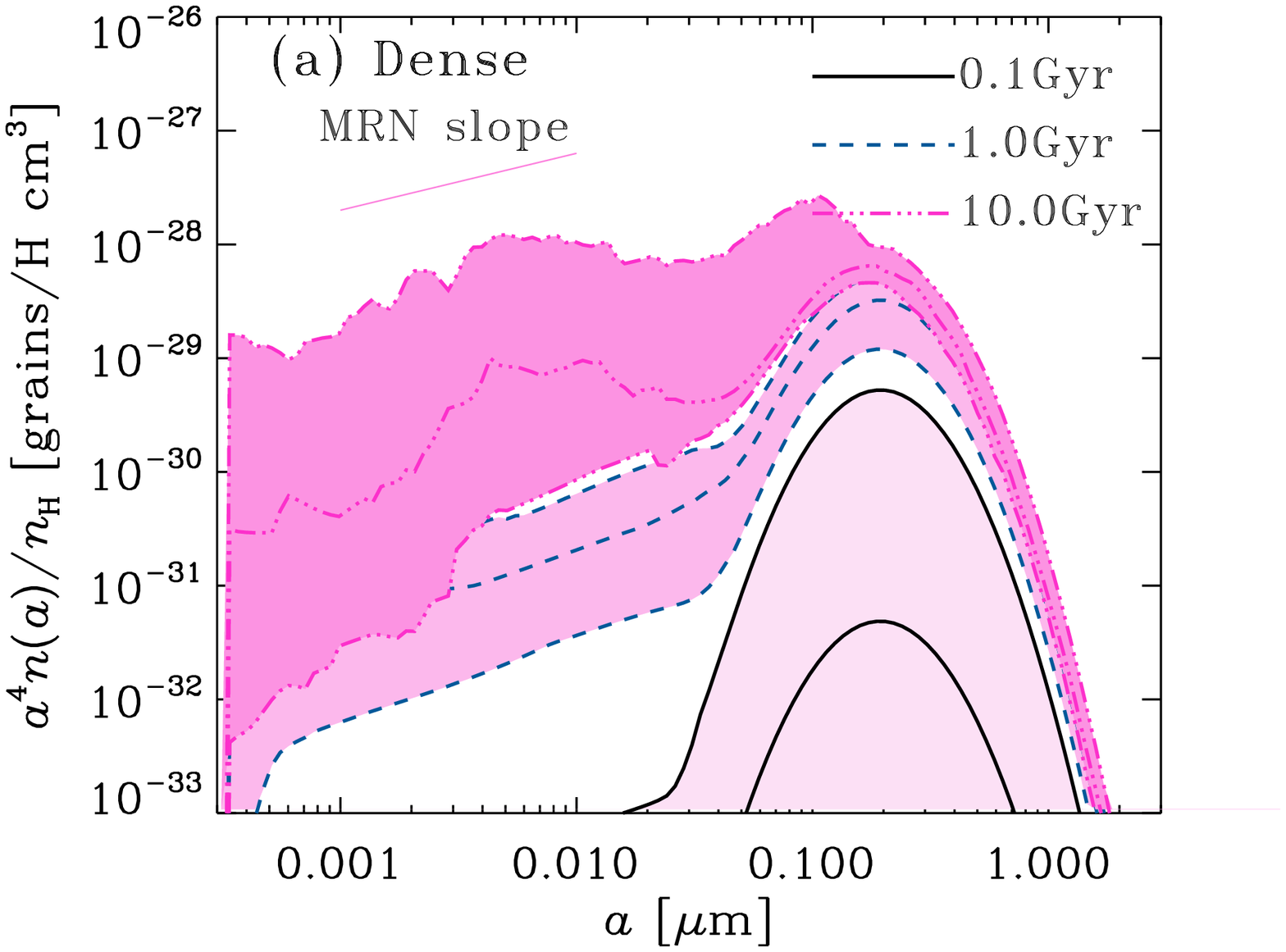}
 \includegraphics[width=1\columnwidth]{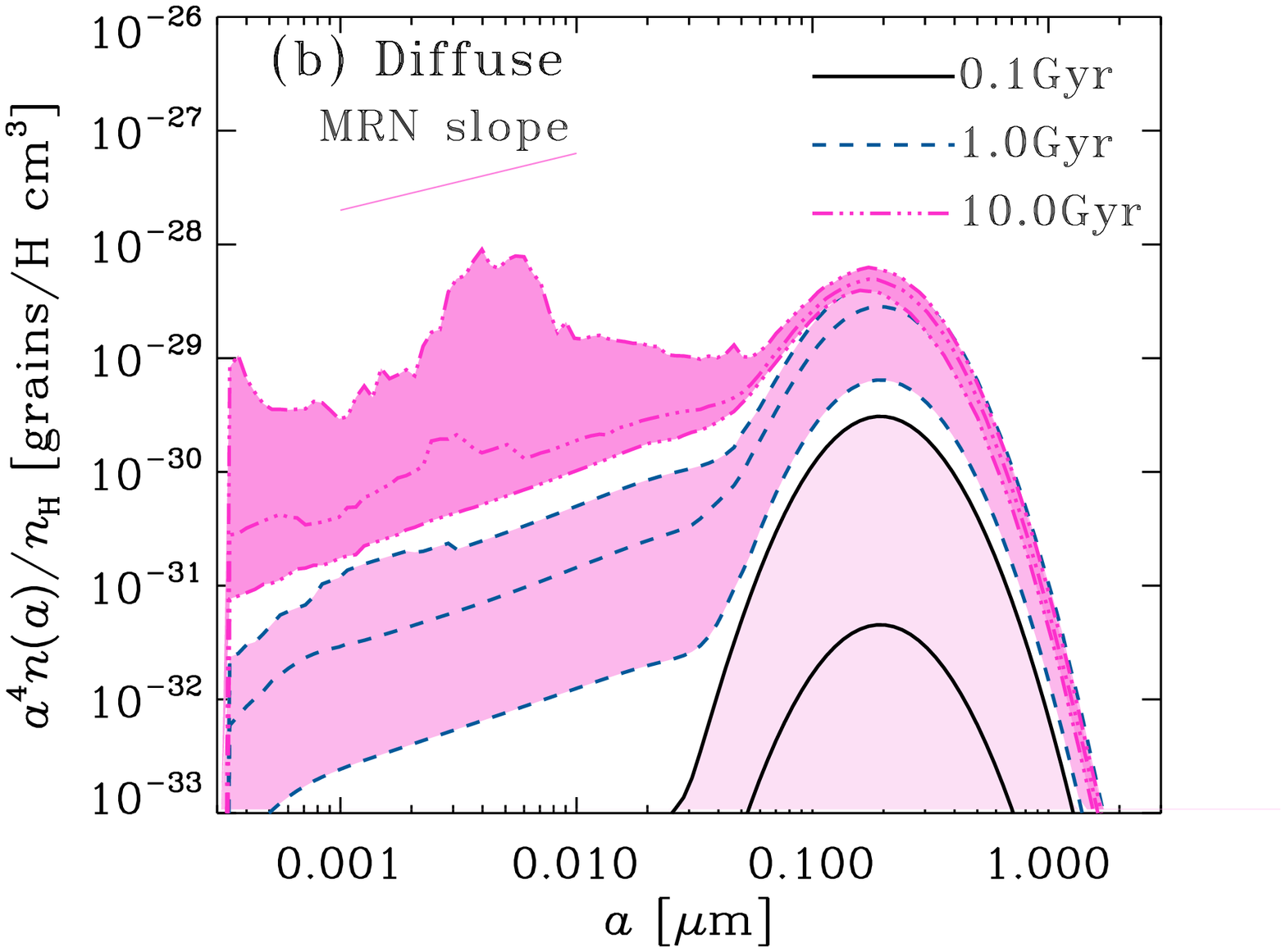}
 \caption{Variation of grain size distribution among the particles
 at $t=0.1$, 1 and 10 Gyr (solid, dashed and triple-dot--dashed lines,
 respectively) in (a) the dense sample and (b) the
 diffuse sample. For each epoch, the middle line is the median
 and the boundaries of the shaded region shows 25th and 75th percentiles
 at each grain radius.
 For $t=0.1$ Gyr, the lower solid line is not shown because it is located below
 the range of the vertical axis.}
 \label{fig:size_ev_statis}
\end{figure*}

To present the statistics property of the grain size distributions in various epochs
($t=0.1$, 1, and 10 Gyr), we show the median and 25th and 75th percentiles
in Fig.\ \ref{fig:size_ev_statis}
(i.e.\ half of the particles in each sample in located in the shaded region).
%%The statistical supports the behaviour of the sampled particles above
%%in Figs.\ \ref{fig:size_ev_dense} and \ref{fig:size_ev_diffuse}.
The grain size distributions are similar between the dense and diffuse samples
at $t=0.1$ and 1 Gyr. This is naturally expected since the physical
states at such young ages do not correlate with those at
$t=10$ Gyr. At $t=10$ Gyr, however, we do see a larger scatter in the dense sample
than in the diffuse sample because of recent dust growth (accretion and coagulation)
in the dense phase. As mentioned above, accretion has a dramatic effect on the
small grain abundance, while coagulation tends to deplete small grains and to
create large grains. Thus, the interplay between accretion and coagulation produces
a large scatter in the grain size distribution both at large and small grain radii
depending on the duration of dust residence in cold and dense gas.
In summary, the current ISM phase affects the grain size distribution
in such a way that the grain size distribution in the dense gas phase tends to have
a larger variation.

The resulting grain size distributions are also compared with the MRN
slope in Fig.\ \ref{fig:size_ev_statis}. Although the dispersion is large,
the grain size distribution at $t=10$ Gyr is broadly consistent with the MRN
slope in both dense and diffuse samples. The grain size distributions at
$t=0.1$ and 1 Gyr are dominated by large grains.
Therefore, the interstellar processing produces MRN-like grain
size distributions at $t\gtrsim 3$ Gyr, which is
comparable to the age of the Milky Way.

\section{Discussion}\label{sec:discussion}

In the above, we have implemented our evolution model of grain size
distribution in a hydrodynamical evolution model of the ISM by post-processing.
We have confirmed that the new framework reproduces the
evolutionary trend of grain size distribution shown in previous work;
that is, the evolution from grain size distributions dominated by large
grains to MRN-like ones as a result of interstellar processing
(especially interplay between accretion and coagulation).
At the same time, we have also shown that there is a large dispersion in the grain
size distributions especially in the later epoch ($t\gtrsim 3$ Gyr).

Although the implementation of our grain size evolution model is
{a successful first step}, there are some observational and computational issues.
Prediction of observable quantities is important for testing the model.
As done by previous studies such as \citet{Hou:2016aa,Hou:2017aa}, we calculate
extinction curves for the purpose of comparison with observations.
The computational issue, on the other hand, is related to the capability
required to simultaneously solve hydrodynamics
and grain size distribution. We discuss these issues in
the following subsections.

\subsection{Extinction curves}\label{subsec:extinction}

Extinction curves have been useful in constraining the grain size distribution
\citep[e.g.][]{Weingartner:2001aa}. Here we calculate extinction curves
based on the grain size distributions obtained
in Section \ref{sec:result}.
The extinction at wavelength
$\lambda$ in units of magnitude ($A_\lambda$)
%%normalized to the column density of hydrogen nuclei ($N_\mathrm{H}$)
is written as
\begin{align}
%%\frac{A_\lambda}{N_\mathrm{H}}
A_\lambda =2.5\log\mathrm{e}\sum_i\int_0^\infty
%%\frac{n_i(a)}{n_\mathrm{H}}
n_i(a)\upi a^2Q_\mathrm{ext}(a,\,\lambda ),
\end{align}
where $Q_\mathrm{ext}(a,\,\lambda )$ is the extinction efficiency
factor, which is evaluated by using the Mie theory \citep{Bohren:1983aa}
with the same optical constants for
silicate and carbonaceous dust (graphite) as in
\citet{Weingartner:2001aa}. The subscript $i$ means the grain composition
(silicate and carbonaceous dust).
For the first step, the mass fractions of silicate
and carbonaceous
dust are fixed to 0.54 and 0.46, respectively,
(number ratio 0.43 : 0.57) with the same grain size distribution
\citep{Hirashita:2009ab}.
Because this fraction is
valid only for the Milky Way, we concentrate on the comparison with the
Milky Way extinction curve below. For other extinction curves, we need
further elaboration on the separate treatment of silicate and carbonaceous dust
\citep{Bekki:2015aa,Hou:2016aa}.
{We also note that there are other grain materials that could explain the
Milky Way extinction curve \citep{Zubko:2004aa,Jones:2013aa}.
For simplicity, we concentrate on the above silicate--graphite model in this
paper.}

\begin{figure*}
 \includegraphics[width=1\columnwidth]{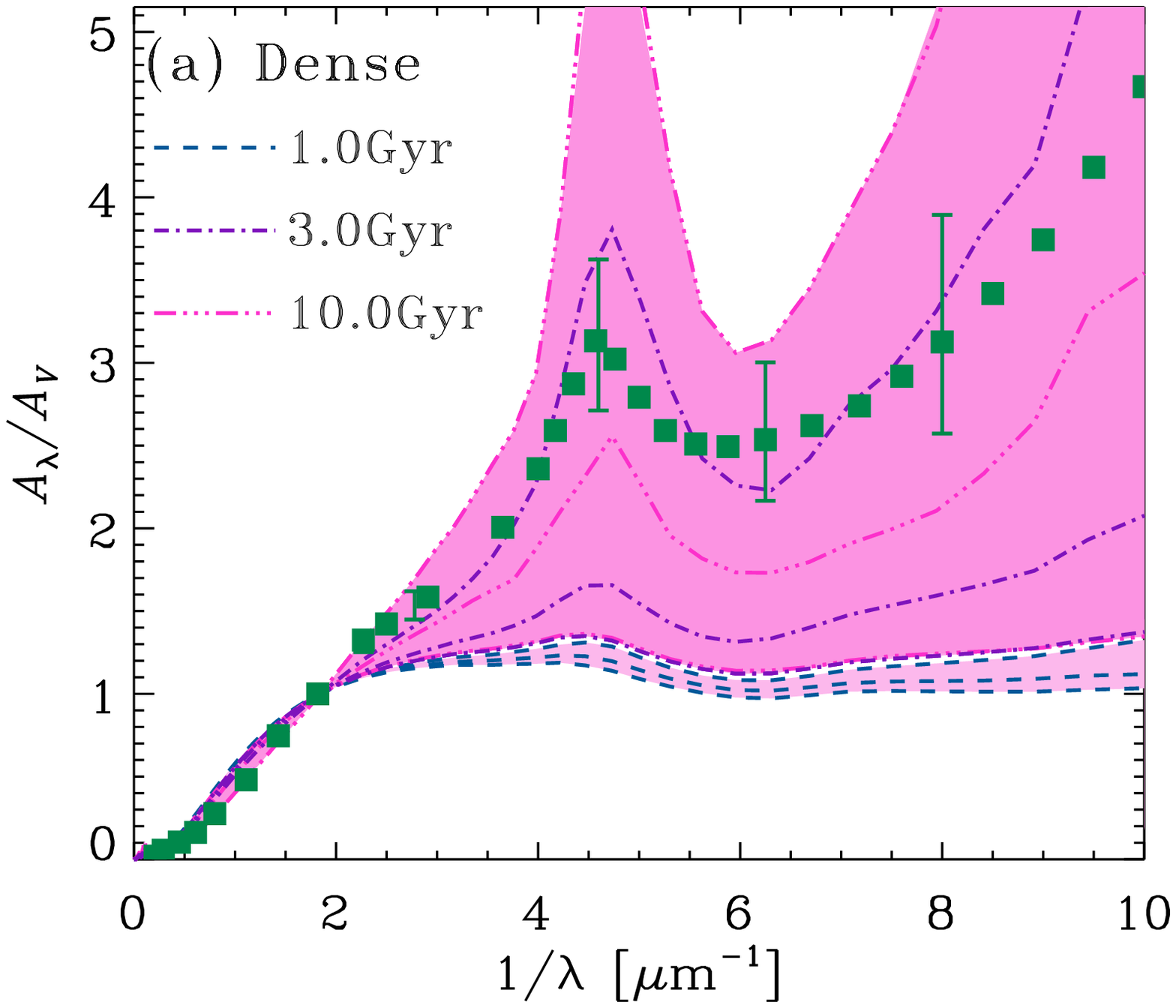}
 \includegraphics[width=1\columnwidth]{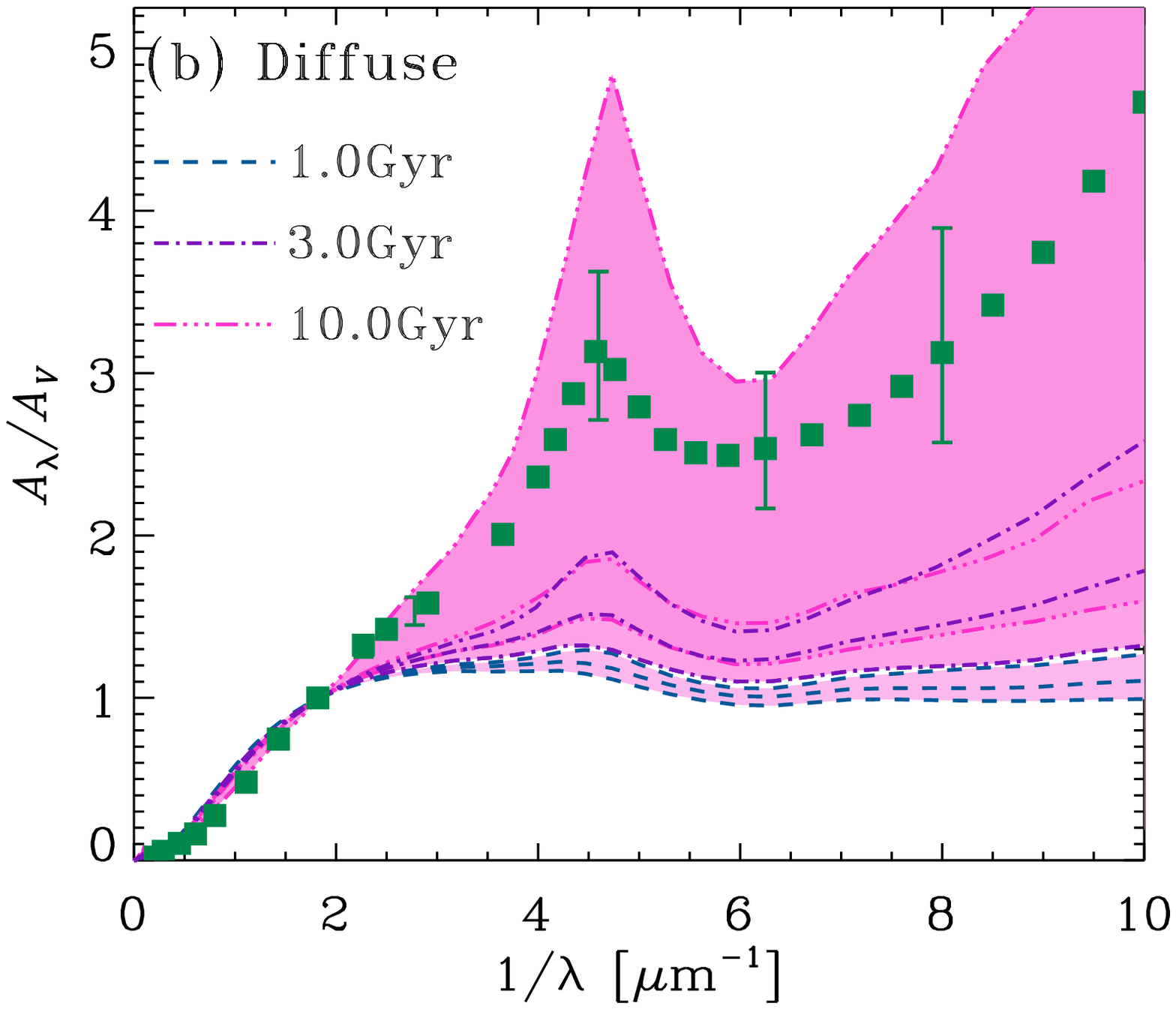}
 \caption{Variation of extinction curves among the gas particles
 at $t=1$, 3 and 10 Gyr (dashed, dot--dashed and triple-dot--dashed lines,
 respectively) for (a) the dense sample and (b) the
 diffuse sample. For each epoch, the middle line is the median at each grain
 size bin and the boundaries of the shaded region shows 25th and 75th percentiles
 at each wavelength.
 Most of the shaded region for $t=3$ Gyr is located behind that for $t=10$ Gyr.
 For comparison, we also show the mean observed extinction curve in the Milky Way
 (squares; \citealt{Pei:1992aa}) and the dispersion among various lines of sight
 (bars; \citealt{Fitzpatrick:2007aa,Nozawa:2013aa}).}
 \label{fig:ext_statis}
\end{figure*}

In Fig.\ \ref{fig:ext_statis}, we show the extinction curves
at $t=1$, 3, and 10~Gyr (the extinction curves
at $t<1$ Gyr are as flat as the ones at $t=1$ Gyr).
To concentrate on the extinction curve shape, we normalize all the
extinction curves to the $V$-band value.
We show the median and 25th and 75th percentiles at each age.
At $t\lesssim 1$ Gyr, the extinction curves are flat, reflecting the
grain size distribution dominated by the large grains produced by stars.
As the abundance of small grains increases, the extinction curve
becomes steeper with more prominent 2175~\AA\ bump created by
small carbonaceous grains. Because of the large variation in the
small grain abundance relative to the large grain abundance,
the extinction curve shape
has a large variation at UV wavelengths at $t=10$ Gyr. As we observe
in Fig.\ \ref{fig:size_ev_statis}, the variety in the grain size
distribution is large at grain radii $a\lesssim 0.05~\micron$
at $t=10$ Gyr,
which affects the extinction at $\lambda\lesssim 2\pi a\sim 0.3~\micron$.

We also observe some difference between the dense and cold gas
in the extinction curves shown in Fig.~\ref{fig:ext_statis}.
As clarified in Fig.\ \ref{fig:size_ev_statis}, the scatter in the
grain size distribution is larger in the dense gas than in the diffuse gas
at $t=10$ Gyr.
As a consequence, the variation in the extinction curves is larger
in the dense gas than in the diffuse gas.
For comparison, we show the observed Milky Way extinction curve and its
variation in various lines of sight \citep{Pei:1992aa,Fitzpatrick:2007aa,Nozawa:2013aa}.
It seems that the extinction curves in the
simulation have a larger variety than the observed scatter. 
However, it is likely that the scatter of extinction curves are significantly
overestimated compared with the observed variance, because
an observational extinction curve represents the averaged
extinction curve in a line of sight,
which could contain gas with a variety of physical states.
Therefore, the dispersion in the simulated extinction curves
for the individual gas particles can be taken as an upper limit for the
actually observed variation in extinction curves.

As we observe in Fig.\ \ref{fig:size_ev_statis}, the median of the extinction
curves at $t=10$ Gyr is steeper in the dense gas than in the
diffuse gas. In contrast, \citet{Hou:2017aa}, who represented
the entire grain size range by two sizes
(two-size approximation proposed by \citealt{Hirashita:2015aa})
in their theoretical model,
showed that the extinction curves at $t=10$ Gyr is flatter in the dense ISM
than in the diffuse ISM because of more efficient coagulation.
The variation in the observed UV extinction curve slope
\citep{Fitzpatrick:2007aa} can also be
interpreted as extinction curves being flatter in the denser ISM
\citep{Hirashita:2014ab,Hou:2017aa}. This implies that our model
overproduces the relative abundance of small grains to large grains
in the dense gas. We make some efforts of resolving this issue
in the next subsection.

The above simple model of silicate--graphite mixture may not be
applied to extinction curves which do not have a clear 2175 \AA\ bump.
\citet{Pei:1992aa} and \citet{Weingartner:2001aa} argued that the
extinction curves in the Large and Small Magellanic Clouds (LMC and
SMC), which have a weaker or no 2175 \AA\ bump,
could be reproduced by decreasing the fraction of graphite.
However, \citet{Hou:2016aa}, using their dust evolution models,
showed that small carbonaceous grains are inevitably produced
as a result of interstellar processing. Thus, they suggested that the
extinction curves without a prominent bump can be produced
either by selectively destroying small carbonaceous grains by
supernova destruction or by introducing amorphous carbon
instead of graphite. \citet{Bekki:2015aa} proposed that small carbonaceous
grains are preferentially lost through radiation-driven wind during the
most recent starburst in the SMC.
\citet{Nozawa:2015aa} argued that the bumpless extinction curve in
a high-redshift quasar can also be explained by using amorphous
carbon instead of graphite. Therefore, it seems that reproducing
the LMC and SMC extinction curves contains problems not related to the
evolution of grain size distribution. Further detailed modeling of carbonaceous
dust is left for the future work, which could address the origin of the
LMC and SMC extinction curves.

\subsection{Necessity of further tuning of the subgrid models?}
\label{subsec:dense_subgrid}

From the above results, there are two issues to resolve.
(i) As we observe in Fig.\ \ref{fig:ext_statis}, the median extinction curve
at $t=10$ Gyr is flatter than the observed
Milky Way extinction curve, which indicates that small grains
are underproduced in our model.
(ii) As we observe in the same figure, the extinction curves in the dense
gas are on average steeper than those in the diffuse gas, while
the opposite trend (i.e.\ flatter extinction curves in denser gas) is
indicated observationally.
These two points indicate that
we need to make the small grain production more efficient
(in order to make the overall extinction curve shape steeper) but
that we require more conversion from small grains to large grains
in the dense gas (in order to make the extinction curves in the
dense gas flatter than those in the diffuse gas).

There is a possible solution that could resolve both of the above issues
simultaneously; that is, making dust growth processes in the dense ISM
(accretion and coagulation) more efficient.
Since the small grain production at the late evolutionary stage is dominated
by accretion, increasing the accretion efficiency would help to increase the
overall small grain abundance. At the same time, by increasing the
coagulation efficiency, small grains could be more depleted in the dense gas than
in the diffuse gas. Since both processes are regulated by the subgrid model
(recall that our simulation is not capable of resolving dense clouds hosting
accretion and coagulation; see Section \ref{subsec:post-process}),
we tune the subgrid model in this subsection.

To make accretion and coagulation more efficient, we increase the density
of subgrid dense clouds to $n_\mathrm{H,dense}=10^4$ cm$^{-3}$
(with other parameters fixed). This subgrid model is referred to as the
dense subgrid model.
Recall that we originally assumed
$n_\mathrm{H,dense}=10^3$ cm$^{-3}$ (Section \ref{subsec:post-process};
the original subgrid model is referred to as the standard subgrid model).
\citet{Zhukovska:2016aa} proposed that the efficiency of accretion is
enhanced by the Coulomb focusing for small grains even in the medium
whose density is less than $10^3$ cm$^{-3}$
\citep[see also][]{Zhukovska:2018aa}. However, in our simulation,
we are not able to address the enhancing mechanism of dust growth by
accretion since the regions where accretion occurs are not spatially resolved.
In other words, the essential assumption here is that the accretion efficiency
is enhanced by a certain mechanism; thus, it does not matter if the
accretion is enhanced by the increase of gas density or the Coulomb focusing
(or any other mechanism).

\begin{figure*}
 \includegraphics[width=1\columnwidth]{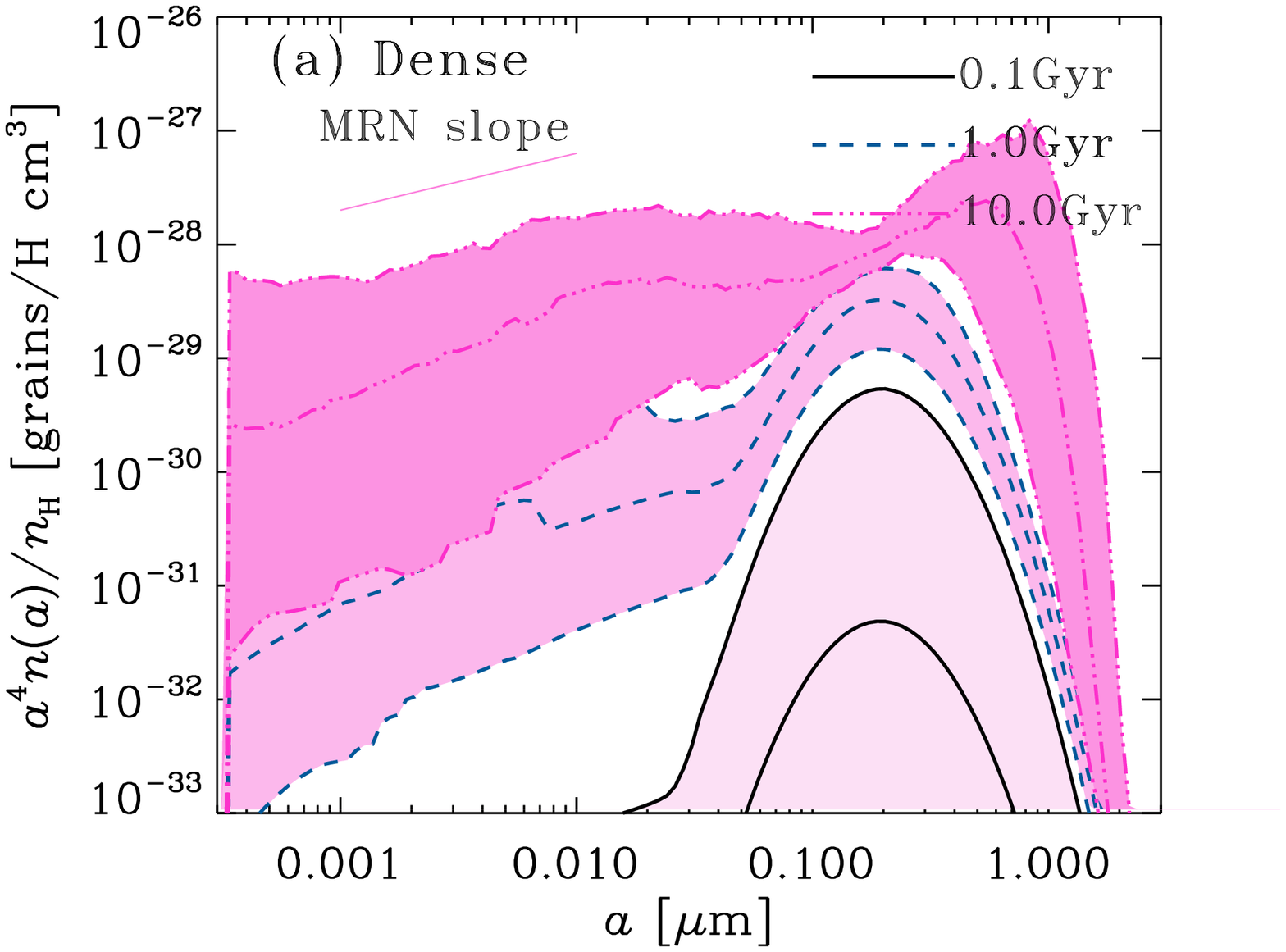}
 \includegraphics[width=1\columnwidth]{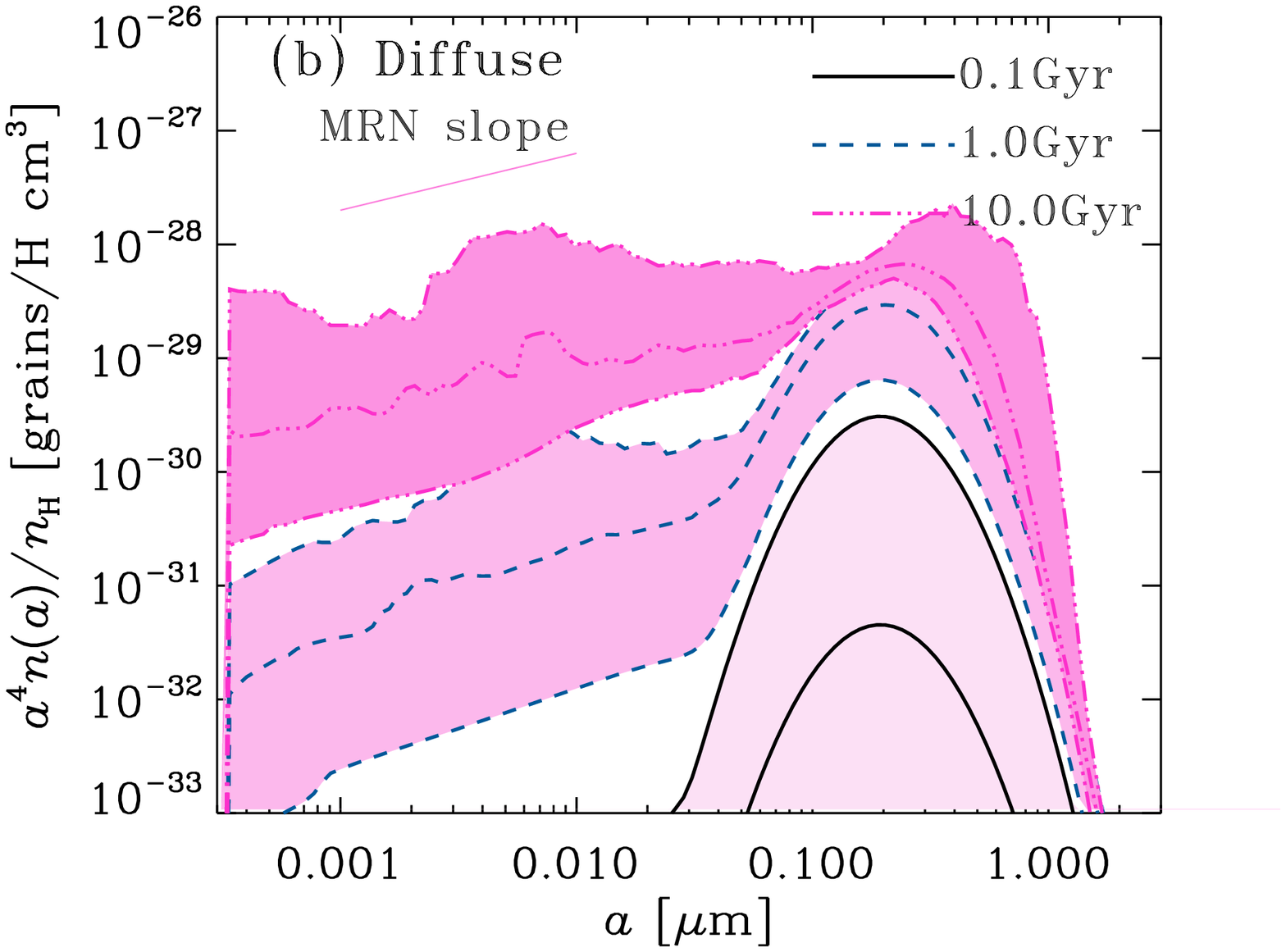}
 \caption{Same as Fig.\ \ref{fig:size_ev_statis_sub1e4} but for the dense
 subgrid model (Section \ref{subsec:dense_subgrid}).}
 \label{fig:size_ev_statis_sub1e4}
\end{figure*}

In Fig.\ \ref{fig:size_ev_statis_sub1e4}, we show the mean grain size
distributions at $t=0.1$, 1, and 10 Gyr with 25th and 75th percentiles
(this figure is to be compared with Fig.\ \ref{fig:size_ev_statis}) for
the dense subgrid model.
Coagulation produces grains as large as $a\sim 1~\micron$,
especially in the dense ISM. As expected, the abundance of small grains
is larger in the dense subgrid model than in the standard subgrid model
at $t=10$ Gyr because of the enhanced accretion rate.

\begin{figure*}
 \includegraphics[width=1\columnwidth]{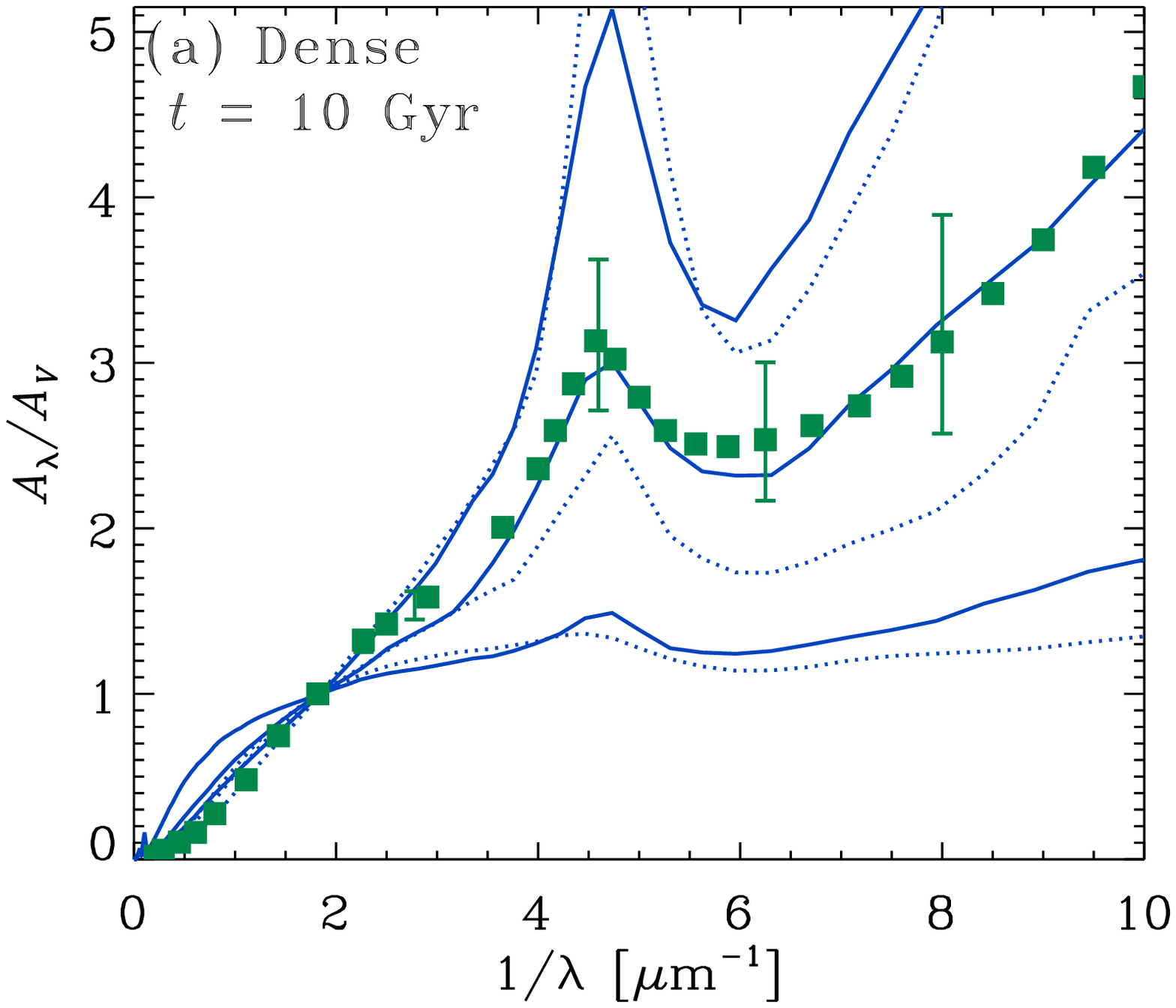}
 \includegraphics[width=1\columnwidth]{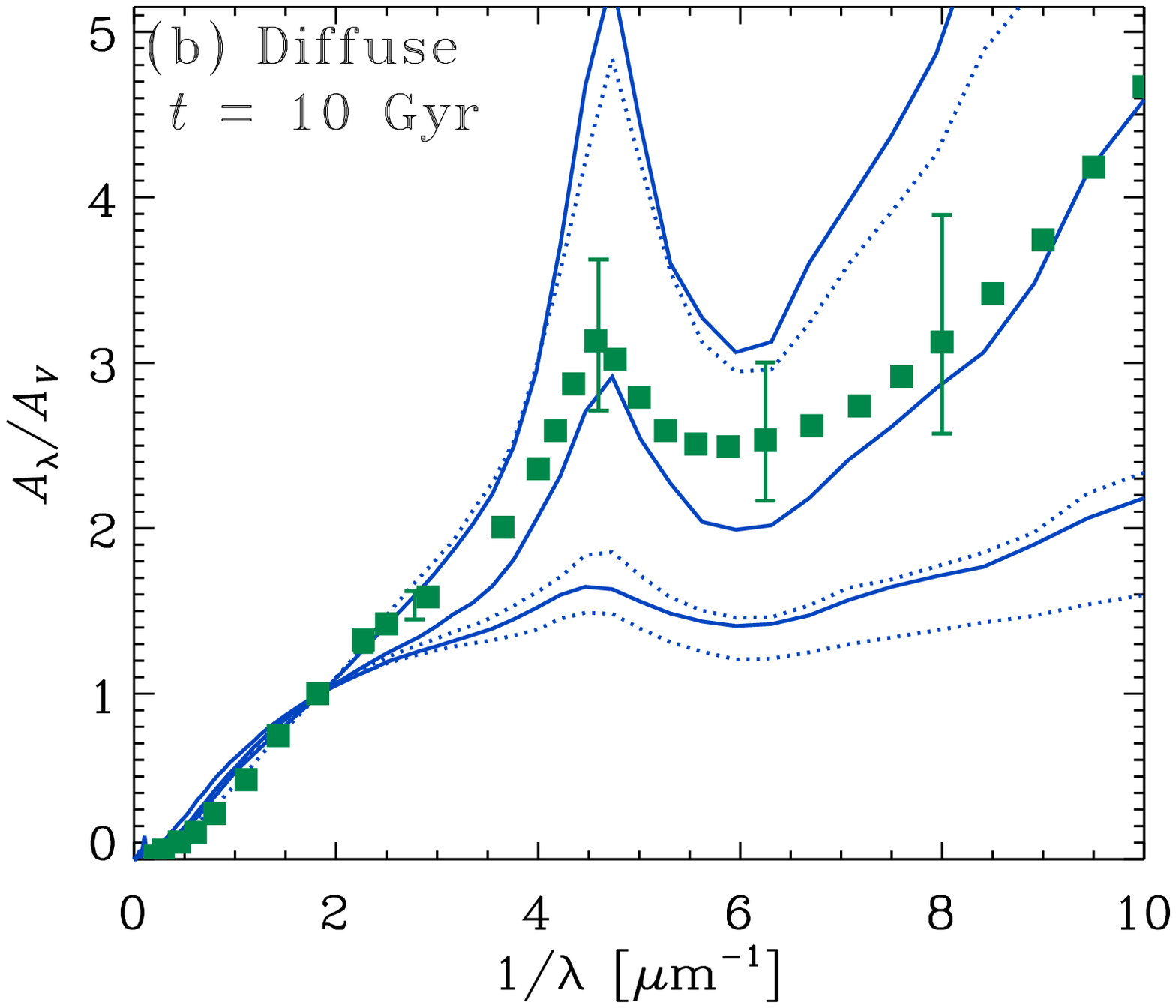}
 \caption{Variation of extinction curves at $t=10$ Gyr. The solid and dotted
 lines show the results for the dense subgrid model
 (Section \ref{subsec:dense_subgrid}) and the standard subgrid model (i.e.\ the
 same model as shown in Fig.\ \ref{fig:ext_statis}), respectively. The middle line is the median
 while the upper and lower lines show 25th and 75th percentiles, respectively.}
 \label{fig:ext_statis_sub1e4}
\end{figure*}

In Fig.\ \ref{fig:ext_statis_sub1e4}, we compare the extinction curves
in the two subgrid models. We only show the results at $t=10$ Gyr,
when the difference between the two subgrid models is the largest. Although the
diversity in the extinction curves is not very different between the
two subgrid models,
the median becomes significantly steeper in the dense subgrid
model than in the standard subgrid model because of the higher
small-grain abundance. Moreover, it is remarkable that the median extinction
curves are consistent with the observed Milky Way extinction curve both in the
dense and diffuse gas. We also observe that the extinction curves are not
significantly different between the diffuse and dense gas in the dense subgrid model.
This means that
we still fail to reproduce the observational trend of flatter extinction curves in
the denser ISM. Further enhancement of coagulation might be necessary,
but it may crete dust grains exceeding $a=1~\micron$. Such large dust grains
are not consistent with observed extinction curves (e.g.\ MRN).
Probably, further fine-tuning of coagulation should include suppression of
coagulation at $a\gtrsim 1~\micron$. We do not further fine-tune the
model in this paper because it is not physically meaningful to complicate the
subgrid model further without strong physical motivation.

It is worth emphasizing that the dense subgrid model -- a simple modification
of the subgrid model -- reproduces the Milky Way extinction curve. Therefore,
our model, combined with the hydrodynamical evolution
of a Milky-Way-like disc galaxy, is a useful tool, based on which we
are able to investigate the evolution of grain size distribution in the entire history of
galaxy evolution.

\subsection{Effects of the number of grain-radius bins}

Calculation of grain size distribution is computing-resource-consuming
when implemented in a hydrodynamical simulation.
Thus, it is interesting to investigate the effect of a reduced grain
radius bin number on the calculated grain size distribution.
We utilize the one-zone model
used in Section \ref{sec:application} for this test to trace a representative
grain size distribution in a galaxy. \citet{McKinnon:2018aa} performed a
test calculation for an isolated disc galaxy and adopted a bin number of
$N=16$ {(although we note that they developed a second-order scheme, while
ours is a first-order one)}. We compare the results with $N=16$ and those with
$N=128$ (the number of bins adopted in all the above calculations) {using
the scheme developed in this paper}.

\begin{figure*}
 \includegraphics[width=1\columnwidth]{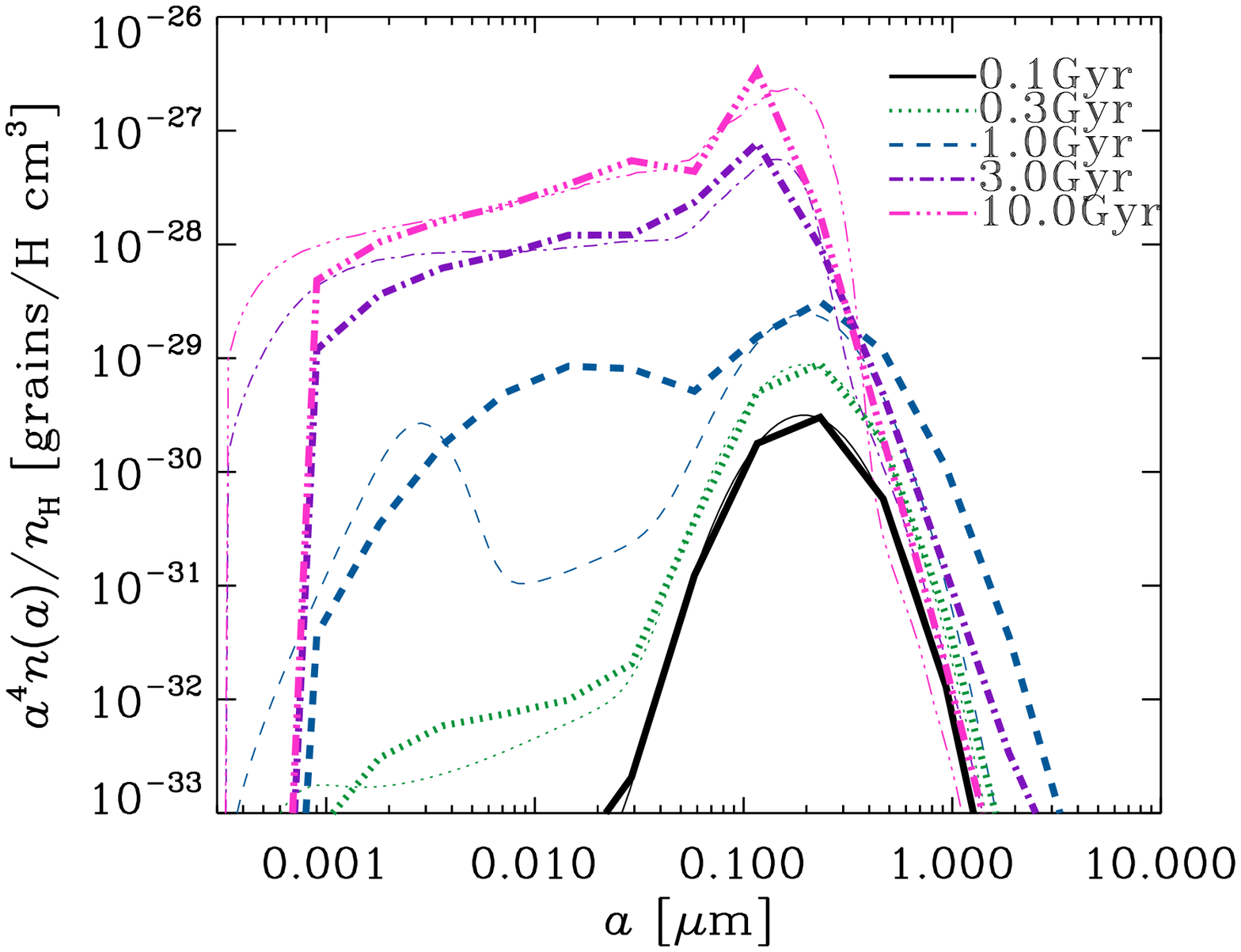}
 \includegraphics[width=1\columnwidth]{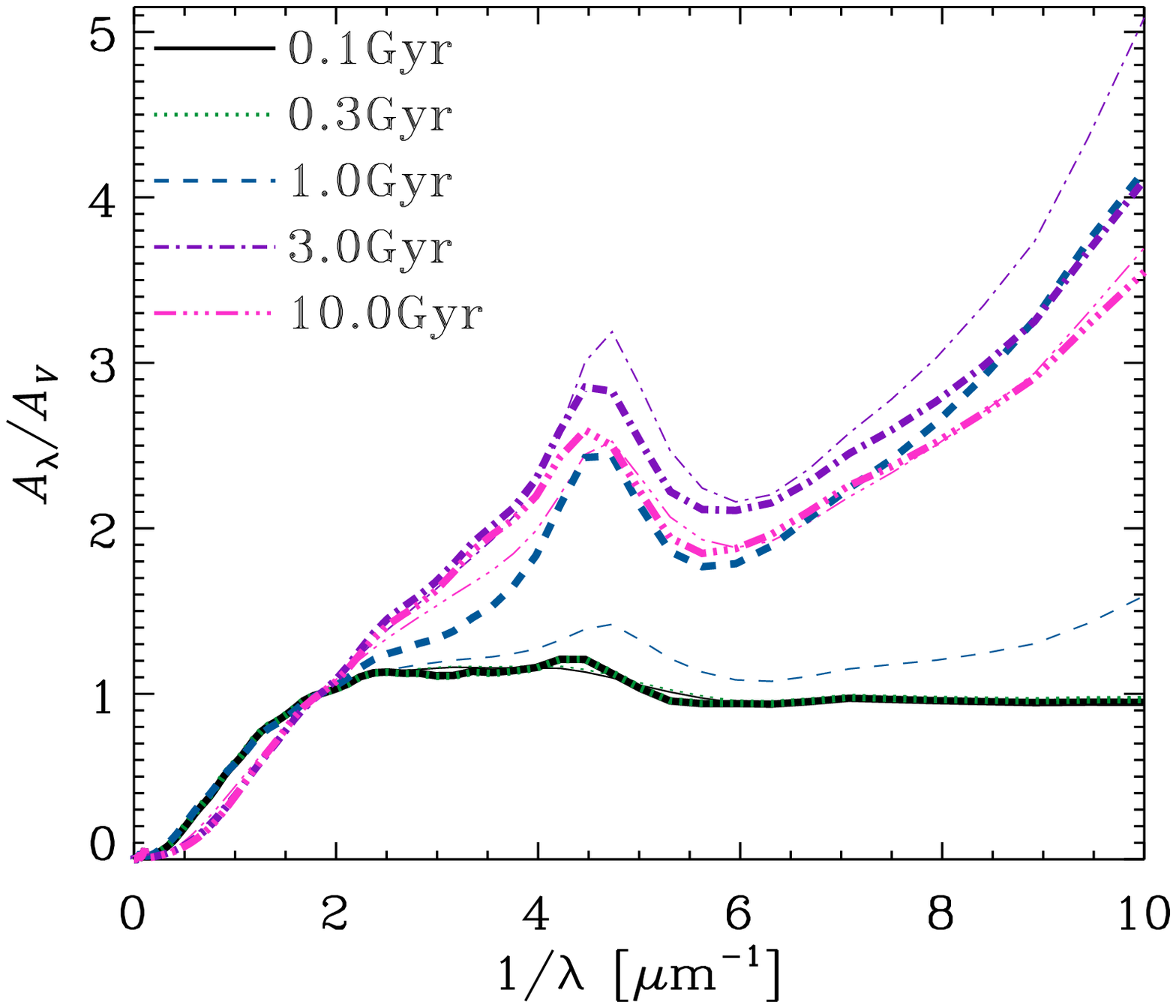}
 \caption{Comparison between the calculations with different grain radius resolutions
 ($N=16$ and 128).
 The thick and thin lines show the calculations with $N=16$ and 128, respectively.
 The solid, dotted, dashed, dot--dashed, triple-dot--dashed
 lines show the results at $t=0.1$, 0.3, 1, 3, and 10 Gyr, respectively.
 \textit{Left}: Evolution of grain size distribution
 for the dense one zone model (i.e.\ the same model as presented in
 Section \ref{sec:application} and Fig.\ \ref{fig:onezone}b). \textit{Right}:
 Extinction curves corresponding to the size distributions in the left panel.}
 \label{fig:resolution}
\end{figure*}

In Fig.\ \ref{fig:resolution}, we show the grain size distribution at
various ages for $N=16$ and 128. We observe that the lower-resolution
run broadly captures the shape of grain size distribution. However, there
is a large difference in the bump structure at $t=1$~Gyr in terms of both the
width and the peak position. This is caused by the numerical diffusion
which is unavoidable in our simple discretization method (especially for
accretion, which is treated by an advection equation in the grain radius
space; see Appendix \ref{app:discrete}).
%%The bump created by
%%accretion around $a\sim 0.003~\micron$ is broadened at $t=1$ Gyr.
Since the evolution of grain size distribution occurs quickly at
$t\sim 1$ Gyr, the shape of the grain size distribution is sensitive to
the slight change in the grain size distribution.
However, at $t=3$ and 10 Gyr, when the
evolution of grain size distribution becomes more moderate than at
$t\sim 1$ Gyr, the low-resolultion run produces
similar grain size distributions to the high-resolution run.

In order to show the effect of grain radius resolution on the prediction of dust optical
properties, we compare the calculated extinction curves in Fig.\ \ref{fig:resolution}.
We find that both resolutions predict similar extinction curve shapes within
$\sim 10$ per cent of difference
at the early ($t\leq 0.3$ Gyr) and late ($t\sim 10$ Gyr) phases.
However, the extinction curve strongly depends on
the grain radius resolution at $t\sim 1$ Gyr as a result of the different
bump shapes in the grain size distribution. Therefore, in the epoch when the
small grain abundance rapidly evolves by accretion, the extinction curve shape
is sensitive to the grain radius resolution.

In summary, the grain size distribution and the extinction curve are
reasonably reproduced with
a low grain radius resolution ($N\sim 16$) except in the phase of rapid dust growth
by accretion. When the grain size distribution is strongly modified by accretion,
the calculated grain size distribution and extinction curve strongly depend on the
grain size resolution. However, in spatially resolved hydrodynamical simulations,
grain growth by accretion does not occur simultaneously in all places; thus,
a strong bump at a certain grain size is not expected if we average the grain
size distribution over the entire galaxy or over a certain area of interest.
Therefore, we expect that a low grain radius resolution with $N\sim 16$
is still useful to
capture the overall shapes of grain size distribution and extinction curve,
considering that calculation with a larger $N$ is computationally expensive.

\subsection{Future prospect}\label{subsec:prospect}

In this paper, we post-processed the hydrodynamical simulation
of an isolated galaxy. However, if dust
grains affect the gas dynamics and/or chemical properties
\citep{Bekki:2015ab}, it is essential to solve hydrodynamics and
dust evolution simultaneously. Moreover, once we implement the
grain size evolution in a hydrodynamical simulation, we naturally
predict, for example, the spatial variation of grain size distribution.
Therefore, it is desirable to include the evolutionary scheme of grain
size distribution developed in this paper in a hydrodynamical simulation.

\citet{McKinnon:2018aa} have already implemented the evolution of
grain size distribution in a hydrodynamical simulation of an isolated
galaxy. Although some important processes in galaxy evolution (especially
stellar feedback) are still
to be included, they have succeeded in showing that the grain size distribution
is strongly influenced by interstellar processing such as shattering
and accretion. %%In the preparatory test calculations they provide,
{Their scheme has some advantages over ours. First, they treated
the grains and gas as separate fluids (i.e.\ allowed to move with different
speeds) but coupled through drag. Second, they considered the
variation of the slope of grain size distribution within each single
grain radius bin to obtain a second-order precision.}
%%However, to save calculation time,
%%they eventually assumed that
%%gas and dust are completely coupled (instead of solving the decoupled motion
%%equations for these components) and fixed the slope in each grain size bin.
{In their calculation, though, saving calculation time still
seems to be an issue for galaxy-scale calculations in their section 5
(for example, they used the first-order scheme in the end).}
In this respect, our formulation could be useful because of
some simplifications in the treatments of sputtering, grain velocities, etc.\
(Section \ref{sec:model}).

The evolution model of grain size distribution developed in this paper is
also useful by itself. As shown in Section \ref{sec:application}, the
application to a one-zone model could be useful to investigate in detail
how each process affects the grain size distribution.
Although \citet{Asano:2013aa} already applied their evolution model of
grain size distribution to a one-zone galaxy evolution model, it would
be interesting to expand the application to different types of galaxies
at various redshifts. Since galaxies have a huge diversity in
star formation histories, our model developed in this paper will provide
a useful tool to investigate the grain size distributions in a variety of
galaxies found by sensitive observational facilities such as ALMA.

\section{Conclusion}\label{sec:conclusion}

We consider the evolution model of grain size distribution in the ISM
of a galaxy. To make the implementation in hydrodynamical
simulations easier, we simplify the previous model in
such a way that some model-dependent assumptions on
dust yield, interstellar processing, and grain velocities are
replaced with simpler functional forms. For the first test of
the developed framework, we apply it to
a one-zone chemical evolution model of a galaxy, confirming
that our new model satisfactorily reproduces the evolutionary
trend of grain size distribution shown in previous work (e.g.\ A13): The dust abundance is
dominated by large grains at $t\lesssim 0.3$--1 Gyr, while
the tail toward small grain radii develops by shattering. After that,
once the metallicity becomes high enough for dust growth by accretion
to be efficient enough, accretion creates a bump at small grain
radius ($a\sim 0.001$--0.01 $\micron$) in the grain size distribution.
This bump is pushed toward larger grain radii and smoothed
by coagulation. We also confirm that efficient grain growth by coagulation
and accretion in the dense ISM is essential in
reproducing the so-called MRN grain size distribution, which is appropriate
for the Milky Way dust.

In order to test if our model can be treated together with the
hydrodynamical evolution of the ISM, we
post-process a hydrodynamical simulation of an isolated disc
galaxy using the new grain evolution model.  We sampled $\sim 80$ gas
(SPH) particles for each phase (dense/cold and diffuse/warm phase)
based on the physical state at $t=10$ Gyr.
According to all the
history of physical conditions (gas density, temperature, and metallicity)
as well as the number of SN shocks, we calculate
the evolution of grain size distribution for each gas particle.
As a consequence of this post-processing, we find that, although the overall evolutionary
behaviour of grain size distribution is similar to the results in the above one-zone
model, there is a large variety in the grain size distribution among the gas particles,
especially at $t\gtrsim 1$ Gyr, mainly because dust growth processes
(accretion and coagulation) that dramatically impact the grain size
distribution are sensitive to when and how long the dust is included
in the cold/dense phase.
The dispersion of the grain size distribution at $t=10$ Gyr is larger in the dense gas than
in the diffuse gas, because
of the recent dust growth (accretion and coagulation) in the dense phase.

For an observational test, we calculate extinction curves based on the grain
size distributions computed above.
At $t\lesssim 1$ Gyr, the extinction curves are flat because
the grain size distribution is dominated by large (sub-micron-sized) grains
produced by stars.
To reproduce the Milky Way extinction curve,
the processes in the dense ISM (i.e.\ accretion and coagulation)
play a central role. If these processes are efficient enough as we assumed in
the dense subgrid model in Section \ref{subsec:dense_subgrid},
the median extinction curves are consistent with the Milky Way extinction curve
in both dense and diffuse gas.

Finally, we examine the effect of degraded grain radius resolution,
considering the limitation in computational resources.
We show that a calculation with a small number of grain radius bins ($N\sim 16$)
is still able to capture the overall shape of grain size distribution and
extinction curve except in the phase of rapid dust growth by accretion,
which happens around $t\sim 1$ Gyr in our simulation. However, we expect that, since
all hydrodynamical elements in a galaxy do not coherently experience
this dust growth phase, we still expect that a low grain radius resolution with
$N\sim 16$ is useful when we are interested in averaged dust properties in a certain area.

\section*{Acknowledgements}

%%We are grateful to A. R\'{e}my-Ruyer for providing us with
%%the data for the relation between dust-to-gas ratio and metallicity
%%of nearby galaxies.
We thank K. Nagamine, K.-C. Hou, I. Shimizu, T. Nozawa,
G. Popping, and L. Pagani for useful discussions.
We are grateful to V. Springel for providing us with the original version of the
\textsc{gadget-3} code. Numerical computations were carried out on the Cray XC50 at the Center for Computational Astrophysics, National Astronomical Observatory of Japan.
We also thank the staff in the Theoretical Institute for Advanced Research
in Astrophysics (TIARA) at Academia Sinica Institute of Astronomy and Astrophysics
(ASIAA) for their continuous support.
HH is supported by the Ministry of Science and Technology
grant (MOST 105-2112-M-001-027-MY3 and
MOST 107-2923-M-001-003-MY3).



%\bibliographystyle{mnras}
\bibliographystyle{mnras}
\bibliography{/Users/hirashita/bibdata/hirashita}
%%\bibliography{hirashita}

%%%%%%%%%%%%%%%%% APPENDICES %%%%%%%%%%%%%%%%%%%%%

\appendix

\section{Equations for Sputtering and Accretion}\label{app:derivation}

The common feature for sputtering and accretion is that these processes
conserve the total number of dust grains except at the minimum grain
radius below which the material should be treated as molecules rather
than dust. Therefore, the following continuity equation for the grain size
distribution holds for sputtering and accretion:
\begin{align}
\frac{\upartial n(a,\, t)}{\upartial t}+\frac{\upartial}{\upartial a}[\dot{a}n(a,\, t)] =0.
\end{align}
The grain size distribution $n(a,\, t)$ is related to the mass distribution
function $\rho_\mathrm{d}(m,\, t)$ in the following way (see equation \ref{eq:rho_n}):
\begin{align}
\rho_\mathrm{d}=\frac{1}{3}an.
\end{align}
Using this and $\dot{m}=4\upi a^2s\dot{a}$, and
$\upartial /\upartial m=[1/(4\upi a^2s)]\upartial /\upartial a$, we obtain
\begin{align}
\frac{\upartial\rho_\mathrm{d}(m,\, t)}{\upartial t}
+\frac{\upartial}{\upartial t}[\dot{m}\rho_\mathrm{d}(m,\, t)]
=\dot{a}n=\frac{\dot{m}}{m}\rho_\mathrm{d}(m,\, t).
\end{align}
Furthermore, by introducing a new variable $\mu\equiv \ln m$, we obtain
\begin{align}
\frac{\upartial\rho_\mathrm{d}(\mu ,\, t)}{\upartial t}
+\frac{\upartial}{\upartial\mu}[\dot{\mu}\rho_\mathrm{d}(\mu ,\, t)]=0.
\label{eq:rho_log}
\end{align}
In deriving this equation, we used $\dot{\mu}=\dot{m}/m$
and $\mathrm{d}m /\mathrm{d}\mu =m$.
Equation (\ref{eq:rho_log}) is useful, since we often adopt grids
with a logarithmically equal width in numerical calculations.
It is interesting to point out that $\rho_\mathrm{d}$ is expressed in
a complete conservative form for accretion and sputtering
if we use a logarithmic grid.

\section{Discrete Forms}\label{app:discrete}

For the reader's convenience, we explicitly write the discretized form
for each process.
For numerical calculation, we consider $N$ discrete
grain radii, and denote the lower and upper
bounds of the $i$th ($i=1,\,\cdots ,\, N$) bin as
$a_{i-1}^\mathrm{(b)}$
and $a_i^\mathrm{(b)}$, respectively. We adopt
$a_{i}^\mathrm{(b)}=a_{i-1}^\mathrm{(b)}\delta$,
$a_0^\mathrm{(b)}=a_\mathrm{l}$, and
$a_N^\mathrm{(b)}=a_\mathrm{u}$ with
$\log\delta =(1/N)\log (a_\mathrm{u}/a_\mathrm{l})$.
We represent the grain radius and mass in the $i$th bin
with
$a_i\equiv (a_{i-1}^\mathrm{(b)}+a_i^\mathrm{(b)})/2$
and $m_i\equiv (4\pi /3)a_i^3s$. The
boundary of the mass bin is defined as
$m_i^\mathrm{(b)}\equiv (4\pi /3)[a_i^\mathrm{(b)}]^3s$.
The interval of logarithmic mass grids is denoted as
$\Delta\mu =3\delta$. We also discretize the time and determine
the time-step according to Section \ref{app:timestep}.
%%We use integer indexes $i$ and $n$ to
%%specify the discrete grain radius and time, respectively.
We adopt $N=128$,
$a_\mathrm{l}=3\times 10^{-4}~\micron$,
and $a_\mathrm{u}=10~\micron$. We apply
$n(a_0,\, t)=n(a_N,\, t)=0$ for the boundary
condition.

We denote the value of quantity $Q$ at a discrete
grid as $Q_i^n$, where $i$ and $n$ specify the grain size and time,
respectively.
%%The equation is solved in the discrete grain mass
%%coordinate $m_i$.

\subsection{Sputtering and Accretion}

We obtain the following equation as a
discrete form of equation (\ref{eq:rho_log}) with time and grid
widths of $\Delta t$ and $\Delta\mu$, respectively:
\begin{align}
\frac{\rho_{\mathrm{d},i}^{n+1}-\rho_{\mathrm{d},i}^n}{\Delta t}
+\frac{\dot{\mu}_i^n\rho_{\mathrm{d},i}^n-\dot{\mu}_{i-1}^n\rho_{\mathrm{d},{i-1}}^n}{\Delta\mu}=0
\end{align}
or
\begin{align}
\frac{\rho_{\mathrm{d},i}^{n+1}-\rho_{\mathrm{d},i}^n}{\Delta t}
+\frac{\dot{\mu}_{i+1}^n\rho_{\mathrm{d},i+1}^n-\dot{\mu}_{i}^n\rho_{\mathrm{d},{i}}^n}{\Delta\mu}=0.
\end{align}
To obtain physically reasonable results, it is desirable to adopt
upwind differencing. Therefore, we use the first equation for accretion,
which always increases the grain mass, while we apply the second for
sputtering, which always decreases the grain mass. Accordingly,
we obtain
\begin{align}
\rho_{\mathrm{d},i}^{n+1}=\rho_{\mathrm{d},i}^n-\frac{\Delta t}{\Delta\mu}
(\dot{\mu}_i^n\rho_{\mathrm{d},i}^n-\dot{\mu}_{i-1}^n\rho_{\mathrm{d},{i-1}}^n)
\end{align}
for accretion, and
\begin{align}
\rho_{\mathrm{d},i}^{n+1}=\rho_{\mathrm{d},i}^n-\frac{\Delta t}{\Delta\mu}
(\dot{\mu}_{i+1}^n\rho_{\mathrm{d},i+1}^n-\dot{\mu}_{i}^n\rho_{\mathrm{d},{i}}^n)
\end{align}
for sputtering.

\subsection{Shattering}\label{app:shattering}

The mass density of grains contained in the $i$th
bin, $\tilde{\mathcal{M}}_i^n$ is defined as
$\tilde{\mathcal{M}}_i^n\equiv\rho_i^nm_i\Delta\mu$.
The time evolution of $\tilde{\mathcal{M}}_i$ by
shattering can be written after discretizing equation (\ref{eq:shat}) as
\begin{align}
\frac{\tilde{\mathcal{M}}_i^{n+1}-\tilde{\mathcal{M}}_i^n}{\Delta t}
&=
-m_i\tilde{\mathcal{M}}_i^{n+1}\sum_{\ell =1}^{N}\alpha_{\ell i}
\tilde{\mathcal{M}}_\ell^n\nonumber\\
&+ \sum_{j=1}^{N}
\sum_{\ell =1}^N\alpha_{\ell j}
\tilde{\mathcal{M}}_\ell^n
\tilde{\mathcal{M}}_j^nm_\mathrm{shat}^{\ell j}(i),\label{eq:shat_discrete}
\end{align}
and
\begin{align}
\alpha_{\ell j} =
\frac{\sigma_{\ell j}v_{\ell j}}{m_\ell m_j},
\end{align}
where $m_\mathrm{shat}^{\ell j}(i)$ is the total mass of the shattered
fragments of a grain in the $\ell$th bin that enter the $i$th bin in
the collision with a grain in the $j$th bin, and $\sigma_{\ell j}$ and
$v_{\ell j}$ are, respectively, the grain--grain collisional cross-section
and the relative collision speed between grains in bins $\ell$ and $j$.
The cross-section for shattering is
$\sigma_{\ell j}=\pi (a_\ell +a_j)^2$ (see equation \ref{eq:sigma}).
The total fragment mass in the $i$th bin is determined by
\begin{align}
m_\mathrm{shat}^{\ell j}(i)=\int_i\mu_\mathrm{frag}(m;\, m_\ell ,\, m_j)\mathrm{d}m,
\end{align}
where the fragment mass distribution $\mu_\mathrm{frag}$
is given in equation (\ref{eq:frag}) and the integration is executed in the
radius range of the $i$th bin.
The fragments whose radii are smaller than $a_\mathrm{l}$ are removed
from the calculation.
In equation (\ref{eq:shat_discrete}), the first term on the right-hand
side adopt a semi-implicit method; that is, we evaluate
$\tilde{\mathcal{M}}_i$ at the new time-step ($n+1$).
This stabilizes the calculation. As a consequence, we obtain
\begin{align}
\tilde{\mathcal{M}}_i^{n+1}=\tilde{\mathcal{M}}_i^n/(1+\mathcal{S}_i)
+\Delta t(\mbox{Second Term})/(1+\mathcal{S}_i),\label{eq:shat_implicit}
\end{align}
where
\begin{align}
\mathcal{S}_i\equiv m_i\Delta t\sum_{\ell =1}^N\alpha_{\ell i}\tilde{\mathcal{M}}_\ell^n ,
\end{align}
and ``Second Term'' is the second term on the right-hand side in equation (\ref{eq:shat_discrete}).
After updating the grain size distribution for the first term, we calculate the second term.

The important aspect of shattering is that it conserves the total dust mass
(if we also count the dust grains that have been removed because of $a<a_\mathrm{l}$).
Therefore, we modify the above formulation to guarantee the mass conservation as
follows. We define
\begin{align}
\mathcal{S}_{i,\ell}\equiv m_i\Delta t\alpha_{\ell i}\tilde{\mathcal{M}}_\ell^n ,
\end{align}
so that $\mathcal{S}_i=\sum_{\ell =1}^N\mathcal{S}_{i,\ell}$.
We aim at guaranteeing the mass conservation in each treatment of
grain collisions between a pair of grain radius bins (the $i$th and $\ell$th bins
in the case above). Let us
consider the collision between grains in the $i$th and $\ell$th bins.
We temporarily renew the value in the $i$th bin as
\begin{align}
\tilde{\mathcal{M}}_{i,\ell}^{n+i/N}=\frac{\tilde{\mathcal{M}}_{i,\ell -1}^{n+i/N}}{1+\mathcal{S}_{i,\ell}},
\label{eq:shat_step1}
\end{align}
where we formally write $\tilde{\mathcal{M}}_{i,0}^{n+i/N}=\tilde{\mathcal{M}}_i^{n+(i-1)/N}$
and $\tilde{\mathcal{M}}_{i}^{n+0/N}=\tilde{\mathcal{M}}_i^n$ so that
this equation is valid for $\ell =1$ and $(\ell,\, n)=(1,\, 1)$.
This converges to the first term on the right-hand side of equation (\ref{eq:shat_implicit})
after all the $N$ loops ($\ell =1$, ..., $N$) if $\Delta t\to 0$.
Immediately after equation (\ref{eq:shat_step1}), we update
the values in all bins ($1\leq j\leq N$) as
\begin{align}
\tilde{\mathcal{M}}_{j,\ell}^{n+i/N}=\tilde{\mathcal{M}}_{j,\ell -1}^{n+i/N}+
\frac{\mathcal{S}_{i,\ell}}{1+\mathcal{S}_{i,\ell}}\frac{m_\mathrm{shat}^{i\ell}(j)}{m_i}.
\label{eq:shat_step2}
\end{align}
Noting that $\sum_{j=0}^Nm_\mathrm{shat}^{i\ell}(j)=m_i$, we find that the
total dust mass is conserved if we calculate the pair of equations (\ref{eq:shat_step1})
and (\ref{eq:shat_step2}).
We repeat equations (\ref{eq:shat_step1}) and (\ref{eq:shat_step2}) for
$\ell =1$, ..., $N$ (i.e.\ we consider the collisions with grains in all the grain radius
bins). The results after this loop is denoted as
\begin{align}
\tilde{\mathcal{M}}_i^{n+i/N}=\tilde{\mathcal{M}}^{n+i/N}_{i,N}.\label{eq:shat_step3}
\end{align}
We also repeat the loop for all $i$ ($i=1$, ..., $N$).
we obtain the values for all bins ($i$) at the $(n+1)$th time-step as
\begin{align}
\tilde{\mathcal{M}}_i^{n+1}=\tilde{\mathcal{M}}_i^{n+N/N}.
\end{align}

\subsection{Coagulation}\label{app:coag}

%%The time evolution of $\tilde{\mathcal{M}}_i$ by coagulation
%%is written as
%%\begin{align}
%%\frac{\tilde{\mathcal{M}}_i^{n+1}-\tilde{\mathcal{M}}_i^n}{\Delta t}
%%&=
%%-m_i\tilde{\mathcal{M}}_i^n\sum_{\ell =1}^{N}\alpha_{\ell i}
%%\tilde{\mathcal{M}}_\ell^n\nonumber\\
%%&+ \sum_{j=1}^{N}
%%\sum_{\ell =1}^N\alpha_{\ell j}
%%\tilde{\mathcal{M}}_\ell^n
%%\tilde{\mathcal{M}}_j^nm_\mathrm{coag}^{\ell j}(i),
%%\label{eq:coag_discrete}
%%\end{align}
%%and
%%\begin{align}
%%\alpha_{\ell j} =
%%\frac{\sigma_{\ell j}v_{\ell j}}{m_\ell m_j},
%%\end{align}
%%where  the
We use the same procedure as shattering in calculating coagulation
except that we replace $m_\mathrm{shat}^{\ell j}$ with the
coagulated mass $m_\mathrm{coag}^{\ell j}(i)$, which
is determined as follows:
$m_\mathrm{coag}^{\ell j}(i)=m_\ell$ if
$m_{i-1}^\mathrm{(b)}\leq m_\ell +m_j<
m_i^\mathrm{(b)}$;\footnote{There was a typo in
\citet{Hirashita:2009ab}.}
otherwise $m_\mathrm{coag}^{\ell j}(i)=0$.
The cross-section for the coagulation is
$\sigma_{\ell j}=\pi (a_\ell +a_j)^2$ (see equation \ref{eq:sigma} with $\beta =1$).

\subsection{Time-step}\label{app:timestep}

The time-step interval $\Delta t$ is determined for each step as follows.
For accretion and sputtering, since $\dot{a}$ is constant, the
shortest time-scale of grain size change $a/\dot{a}$ is realized at
the smallest radius bin. Therefore, we adopt
$\epsilon_\Delta\Delta\mu/\dot{\mu}$ at the smallest bin
(with $\epsilon_\Delta =0.3$), denoted as
$\Delta t_\mathrm{acc}=\epsilon_\Delta (\Delta\mu/\dot{\mu})_\mathrm{acc}|_0^n$
and $\Delta t_\mathrm{sput}=\epsilon_\Delta (\Delta\mu/\dot{\mu})_\mathrm{sput}|_0^n$,
for accretion and sputtering, respectively. For shattering and coagulation,
we do not know in advance
at which bin the processes have the shortest time-scale. It is possible
to find $\Delta t$ such that $\Delta t$ is shorter than the inverse of the mass
changing rate in all bins. However, this iterative process is
time-consuming. To avoid the iteration, we introduce the following
collision time-scale, $\tau_\mathrm{coll}$, which can be calculated
in an independent manner from the actual shattering and coagulation
calculations:
\begin{align}
\tau_\mathrm{coll}=5\times 10^7
n_\mathrm{H}\left(\frac{a}{0.1~\micron}\right)
\left(\frac{\mathcal{D}}{0.01}\right)^{-1}\left(
\frac{v_\mathrm{gr}(a)}{10~\mathrm{km~s}^{-1}}\right)^{-1}~\mathrm{yr},
\end{align}
where $\mathcal{D}\equiv \rho_\mathrm{d,tot}/(\mu m_\mathrm{H}n_\mathrm{H})$
is the dust-to-gas ratio.
This collision time-scale is derived under an assumption that
all the grains have a single grain radius $a$ (A17). Since the grains
have a large variety in grain radius, which radius we adopt to estimate
above is not obvious. Thus, we simply estimate $\tau_\mathrm{coll}$
around the medium grain radius, and adopt
$\Delta t=\epsilon_\Delta\tau_\mathrm{coll}$ with $\epsilon_\Delta =0.3$.
We denote the time-step determined in this way as
$\Delta t_\mathrm{shat}=(\epsilon_\Delta\tau_\mathrm{coll})_\mathrm{shat}^n$
and $\Delta t_\mathrm{coag}=(\epsilon_\Delta\tau_\mathrm{coll})_\mathrm{coag}^n$ for
shattering and coagulation, respectively.
Because the collision time-scale is derived under an extreme
assumption of single-sized grains, the time-step set in this manner
usually gives satisfactorily short time-scale.

When implementing the grain size evolution in a hydrodynamical
simulation, we update the grain size distribution at each hydrodynamical
time-step $\Delta t_\mathrm{hydro}$. For each process,
if $\Delta t_p$ ($p$ is the process name `acc', `sput', `shat', or `coag')
is shorter than $\Delta t_\mathrm{hydro}$, we run sub-cycles for the
grain size distribution with a time-step $\Delta t_\mathrm{hydro}/n_\mathrm{sub}$ that satisfies
$\Delta t_p>\Delta t_\mathrm{hydro}/n_\mathrm{sub}$ ($n_\mathrm{sub}$ is the
number of subcycles).

\section{Empirical formula for the grain velocity}\label{app:vel}

We derive the empirical formula for the grain velocity
(equation~\ref{eq:vel}). We basically follow \citet{Ormel:2009aa}.
The grain velocity is determined by grain--gas coupling through
gas drag. The gas drag time-scale, $\tau_\mathrm{dr}$ is estimated
as
\begin{align}
\tau_\mathrm{dr}=\frac{sa}{c_\mathrm{g}\rho_\mathrm{g}},\label{eq:tau_dr}
\end{align}
where $c_\mathrm{g}$ is the sound speed, and $\rho_\mathrm{g}$ is the gas
density. We assume that the
turbulent velocity dispersion on a length scale of $\ell$ is described by
the Kolmogorov spectrum:
\begin{align}
v=v_\mathrm{max}\left(\frac{\ell}{L_\mathrm{max}}\right)^{1/3},
\end{align}
where $v_\mathrm{max}$ is the velocity at the size of the largest eddies,
$L_\mathrm{max}$. With this scaling relation, the turnover time, $\tau_\mathrm{turn}$,
is estimated as
\begin{align}
\tau_\mathrm{turn}\equiv\frac{\ell}{v}=
\frac{L_\mathrm{max}v^2}{v_\mathrm{max}^3}.\label{eq:tau_turn}
\end{align}
A grain with a radius of $a$ is coupled with the turbulence on
the scale $\ell$ such that $\tau_\mathrm{dr}=\tau_\mathrm{turn}$.
This
leads to the following estimate using equations (\ref{eq:tau_dr}) and (\ref{eq:tau_turn}):
\begin{align}
v &= v_\mathrm{max}^{3/2}\left(
\frac{sa}{c_\mathrm{g}\rho_\mathrm{g}L_\mathrm{max}}\right)^{1/2}\nonumber\\
&= \left(\frac{v_\mathrm{max}}{c_\mathrm{g}}\right)^{3/2}\left(
\frac{L_\mathrm{max}}{L_\mathrm{J}}\right)^{-1/2}v_\mathrm{gr,Ormel},
\end{align}
where $L_\mathrm{J}\equiv({\upi c_\mathrm{g}^2/G\rho_\mathrm{g}})^{1/2}/2$
is the Jeans length and
$v_\mathrm{gr,Ormel}\equiv c_\mathrm{g}(sa/\rho_\mathrm{g}L_\mathrm{J})^{1/2}$
is the grain velocity derived by \citet{Ormel:2009aa}, who assumed that
$L_\mathrm{max}=L_\mathrm{J}$ and $v_\mathrm{max}=c_\mathrm{g}$ to obtain
the grain velocity. We also adopt $L_\mathrm{max}=L_\mathrm{J}$ but
relax the condition $v_\mathrm{max}=c_\mathrm{g}$
(we could move both quantities, but we move only $v_\mathrm{max}$ because
they are degenerate).
Using $\mathcal{M}\equiv v_\mathrm{max}/c_\mathrm{g}$, we obtain
equation (\ref{eq:vel}).

% Don't change these lines
\bsp	% typesetting comment
\label{lastpage}
\end{document}